\documentclass[]{aastex631}
\usepackage{rotating}
\usepackage[version=4]{mhchem}

\shorttitle{MALBEC radiative transfer intercomparison initiative}
\shortauthors{}
\graphicspath{{./}{./}}

\begin{document}

\title{Modeling Atmospheric Lines By the Exoplanet Community (MALBEC) version 1.0: A CUISINES radiative transfer intercomparison project}

\correspondingauthor{Geronimo L. Villanueva}
\email{geronimo.l.villanueva@nasa.gov}

\author[0000-0002-2662-5776]{Geronimo L. Villanueva}
\affiliation{NASA Goddard Space Flight Center, Greenbelt, MD, USA.}

\author[0000-0002-5967-9631]{Thomas J.Fauchez}
\affiliation{Integrated Space Science and Technology Institute, Department of Physics, American University, Washington DC}
\affiliation{NASA Goddard Space Flight Center, Greenbelt, MD, USA.}
\affiliation{NASA GSFC Sellers Exoplanet Environments Collaboration}

\author[0000-0002-5060-1993]{Vincent Kofman}
\affiliation{Integrated Space Science and Technology Institute, Department of Physics, American University, Washington DC}
\affiliation{NASA Goddard Space Flight Center, Greenbelt, MD, USA.}

\author[0000-0002-0006-1175]{Eleonora Alei}
\affiliation{ETH Zurich, Institute for Particle Physics \& Astrophysics, Wolfgang-Pauli-Str. 27, 8093 Zurich, Switzerland}
\affiliation{National Center of Competence in Research PlanetS (www.nccr-planets.ch)}

\author[0000-0002-3052-7116]{Elspeth K.H. Lee}
\affiliation{Center for Space and Habitability, University of Bern, Gesellschaftsstrasse 6, CH-3012 Bern, Switzerland}

\author[0000-0003-0475-8479]{Estelle Janin}
\affiliation{School of Earth and Space Exploration, Arizona State University, Tempe, AZ 85281, USA}

\author[0000-0002-9338-8600]{Michael D. Himes}
\affiliation{Planetary Sciences Group, Department of Physics, University of Central Florida, USA.}
\affiliation{NASA Postdoctoral Program Fellow, NASA Goddard Space Flight Center, Greenbelt, MD, USA.}

\author[0000-0002-3555-480X]{J\'er\'emy Leconte}
\affiliation{Laboratoire d'astrophysique de Bordeaux, Univ. Bordeaux, CNRS, B18N, all\'ee Geoffroy Saint-Hilaire, 33615 Pessac, France}

\author[0000-0003-1906-5093]{Michaela Leung}
\affiliation{Department of Earth and Planetary Sciences, University of California, Riverside, California, 92521}

\author[0000-0003-0194-5615]{Sara Faggi}
\affiliation{Integrated Space Science and Technology Institute, Department of Physics, American University, Washington DC}
\affiliation{NASA Goddard Space Flight Center, Greenbelt, MD, USA.}

\author[0000-0002-1624-3360]{Mei Ting Mak}
\affiliation{Physics and Astronomy, University of Exeter, Exeter EX4 4QL, UK}

\author[0000-0001-8832-5288]{Denis E. Sergeev}
\affiliation{Physics and Astronomy, University of Exeter, Exeter EX4 4QL, UK}

\author[0000-0002-3868-2129]{Thea Kozakis}
\affiliation{National Space Institute, Technical University of Denmark, Elektrovej 328, 2800 Kgs. Lyngby, Denmark}

\author[0000-0003-4402-6811]{James Manners}
\affiliation{Met Office, FitzRoy Road, Exeter EX1 3PB, UK}

\author[0000-0001-6707-4563]{Nathan Mayne}
\affiliation{Physics and Astronomy, University of Exeter, Exeter EX4 4QL, UK}

\author[0000-0002-2949-2163]{Edward W. Schwieterman}
\affiliation{Department of Earth and Planetary Sciences, University of California, Riverside, CA, USA}

\author[0000-0002-4884-7150]{Alex R. Howe}
\affiliation{NASA Goddard Space Flight Center, 8800 Greenbelt Rd, Greenbelt, MD 20771, USA}
\affiliation{Center for Research and Exploration in Space Science and Technology, NASA/GSFC, Greenbelt, MD 20771}

\author[0000-0003-1240-6844]{Natasha Batalha}
\affiliation{NASA Ames Research Center, Mountain View, CA 94035, USA}


\begin{abstract}
Radiative transfer (RT) models are critical in the interpretation of exoplanetary spectra, in simulating exoplanet climates and when designing the specifications of future flagship observatories. 
However, most models differ in methodologies and input data, which can lead to significantly different spectra. 
In this paper, we present the experimental protocol of the MALBEC (Modeling Atmospheric Lines By the Exoplanet Community) project. MALBEC is an exoplanet model intercomparison project (exoMIP) that belongs to the CUISINES (Climates Using Interactive Suites of Intercomparisons Nested for Exoplanet Studies) framework which aims to provide the exoplanet community with a large and diverse set of comparison and validation of models. The proposed protocol tests include a large set of initial participating RT models, a broad range of atmospheres (from Hot Jupiters to temperate terrestrials) and several observation geometries, which would allow us to quantify and compare the differences between different RT models used by the exoplanetary community. Two types of tests are proposed: transit spectroscopy and direct imaging modeling, with results from the proposed tests to be published in dedicated follow-up papers. To encourage the community to join this comparison effort and as an example, we present simulation results for one specific transit case (GJ-1214 b), in which we find notable differences in how the various codes handle the discretization of the atmospheres (e.g., sub-layering), the treatment of molecular opacities (e.g., correlated-k, line-by-line) and the default spectroscopic repositories generally used by each model (e.g., HITRAN, HITEMP, ExoMol). 
\end{abstract}


\section{Introduction} \label{sec:intro}

Model Intercomparison Projects (MIPs) are the ideal tool to benchmark models and to track down numerical/algorithm issues and model dependencies. MIPs are widely used in the Earth Science community, with one of the most known MIPs being the Coupled Model Intercomparison Project (CMIP), currently in its sixth version \citep{cmip6}, which compares General Circulation Models (GCMs) in the simulation of Earth's past and current climate and the prediction of its future climate. Such model intercomparisons, which have been running for almost three decades, are playing a crucial role in assessing global climate change, predicted by all models, and in forecasting possible future climate scenarios.

In Earth Sciences, many Radiative Transfer (RT) tool intercomparisons have been performed, such as the RAdiative transfer Model Intercomparison \citep[RAMI,][]{Pinty2001}, whose activity focuses on the benchmarking of Earth forest canopy RT models, or more recently the work by \citet{Zawada_amt2021}, which compared seven different limb-scattering models. Another notable Earth Science RT MIP is the Correlated K-Distribution Model Intercomparison Project (CKDMIP) by \citet{Hogan_Matricardi2020}. This MIP benchmarks line-by-line calculations to evaluate the accuracy of existing correlated k-distribution (CKD) models,  explores how accuracy varies with the number of pseudo-monochromatic calculations, assesses how different choices made when CKD models are generated affect their accuracy, and  generates open-access datasets and software, enabling the development of new gas-optics tools. To date, results from this intercomparison are still being evaluated.

To our knowledge, one of the first RT MIPs to be proposed for the paleoclimate and exoplanet community, was the Palaeoclimate and Terrestrial Exoplanet Radiative Transfer Model Intercomparison Project  \citep[PALAEOTRIP,][]{Goldblatt2017}. No results have been reported from this MIP yet. A few years later, \cite{Barstow2020} compared three radiative transfer codes, TauREx 2 \citep{Waldmann2015a, Waldmann2015b}, NEMESIS \citep{Irwin2008} and CHIMERA \citep[][and reference therein]{MaiLine2019} for hot Jupiter and warm Neptune cases. Good agreements within 20--40~ppm of residual noise were found for the forward models while cross-retrievals were showed to be consistent within $1\sigma$. The same year, \cite{Pincus2020}  developed the Radiative Forcing Model Intercomparison Project (RFMIP) that uses benchmark calculations made with line-by-line models to identify parameterization error in the representation of absorption and emission by greenhouse gases. Among the main findings is that  agreement between line-by-line models is better in the long-wave than in the shortwave where various treatments of the water vapor continuum impact estimates of forcing by carbon dioxide and methane. Other intercomparisons \citep{Baudino2017, Yang2016} have also revealed the importance of benchmarking radiative-transfer and identified notable differences between resulting spectra.

More recently \cite{Barstow2022} developed a retrieval challenge to prepare for the ARIEL mission, with five different exoplanet retrieval codes (ARCiS \citep{Min2020}, NEMESIS \citep{Irwin2008}, Pyrat Bay \citep{Cubillos2021}, TauREx 3 \citep{Al-Refaie2021A3.1} and POSEIDON \citep{MacDonald2017} analyzing the same synthetic dataset of a hot Jupiter (clear and cloudy), a clear warm Neptune and a cloudy super-Earth. Very good agreements among them have been found with the majority of the cases correctly retrieved. \citet{HarringtonEtal2022psjBART1} presented the Bayesian Atmospheric Radiative Transfer (BART) retrieval code alongside a MIP framework called BARTTest, which included a detailed comparison with NEMESIS, CHIMERA, and TauREx from the \citet{Barstow2020} work. As part of the the JWST Early Release Science (ERS) exoplanetary observations, the Transiting Exoplanet Community also explored and compared several RT models \citep{ahrer2022}.

A typical challenge of model intercomparisons is that, as the results converge and become more homogeneous between them, we may reach a situation of high precision (or rather, high robustness) but low accuracy. In that aspect, it is very important to employ models that originate and have foundations from different communities (e.g., Earth remote sensing, Solar System astronomy, astrophysics), where the core elements of the different RT algorithms have been validated and consolidated by comparing to a broad range of datasets. As such, the general philosophy of the CUISINES framework (and therefore of MALBEC) is to achieve high robustness between models while running them using their default configurations, and explore why the results differ and how each model could be improved to reach higher accuracy. The new exoplanet MIP (exoMIP) presented here is called MALBEC (Modeling Atmospheric Lines By the Exoplanet Community). In MALBEC, we intend to go beyond these previous RT model intercomparisons by including a larger set of models ($\sim$~a dozen), a larger and more diverse set of exoplanets (from Hot Jupiters to temperate terrestrials) and several observation geometries (transit spectroscopy, emission spectroscopy and direct imaging). Each model scenario is called a ``cask" within the MALBEC protocol. Currently, only forward modeling test cases are considered in MALBEC, yet we plan to integrate retrievals and noise modeling in future investigations. Comparing simulated spectra across models is a first step, and that allows to determine notable spectroscopic differences, but identifying parametric physical or statistical distances between methods would ultimately require of retrieval analyses that employ different forward models. Also to note is that none of these codes are radiative equilibrium (RE) or radiative convective equilibrium (RCE) models, and the here proposed RT tests operate with prescribed temperature, pressure and chemical structures as input. RE and RCE models will be part of another exoMIP called ``COD ACCRA" for Comparing One-DimentionAl Climates of Convective Radiative Atmospheres, available within the CUISINES framework.

\subsection{Synergy with other model intercomparison projects}
MALBEC is part of the CUISINES model intercomparison work group. CUISINES stands for Climates Using Interactive Suites of Intercomparisons Nested for Exoplanet Studies and provides a framework for different model intercomparison projects, as well as opportunities to find synergies (\textit{pairings}) between different work groups. It currently consists of nine different projects, including chemical, circulation, energy balance, and planet specific models. For more information, see \url{https://nexss.info/cuisines}. The framework started with the TRAPPIST-1 Habitable Atmospheres Intercomparison (THAI) project, which presented a trilogy of GCM studies investigating dry and wet climate for TRAPPIST-1e \citep{turbet_thai,sergeev_thai,fauchez_thai}. CUISINES was inspired by the THAI success, and led to two international workshops and eight more exoMIPs at the time of writing. Three other exoMIPs --- SAMOSA, CAMEMBERT, and FILLET --- have currently published protocol papers describing their efforts:

\begin{itemize}
\item The Sparse Atmospheric Modeling Sample Analysis, or SAMOSA, aims to provide guidance for what type of planets need to be modeled using full 3D general circulation models and where more simple 1D equivalent models are sufficient \citep{HQsamosa}. The parameters investigated are insolation and surface pressure and it compares the outcome scenarios between the ExoCAM 3D GCM and ExoPlaSim 1D models \citep{wolf_exocam_2022,paradise_exoplasim_2022}. 

\item Comparing Atmospheric Models of Extrasolar Mini-Neptunes Building and Envisioning Retrievals and Transits, CAMEMBERT, focuses on inter-comparing the results from eight different GCMs simulating the atmospheres of GJ1214b and K2-18b \citep{christie_camembert_2022}. 

\item Functionality of Ice Line Latitudinal EBM Tenacity, FILLET, studies energy balance models, and in particular looks at climate `end' states using different values for isolation, obliquity, starting temperatures and atmospheric composition \citep{Deitrick2023}. 

\end{itemize}

MALBEC serves several roles within the CUISINES framework: as a down-stream framework to produce spectra and simulate observables from the 3D GCM exoMIPs (THAI, CAMEMBERT, MOCHA), constrain the detectability of outputs from photochemical codes (PIE), and verify the radiative fluxes from energy balance models (FILLET). Central in all of these is the MALBEC configuration file (see below), which provides a standardized and accessible format for both input and output. Namely, ``up-stream'' models that generate profiles of the atmospheric variables (e.g., GCMs, RCE models) can produce output files in the MALBEC format that can then be read by the ``down-stream'' radiative transfer codes. This standardized file provides a bridge for testing the outputs of different models, and is how we ensure the test cases in this protocol paper are interpreted similarly by all codes. Additionally, MALBEC can provide two metrics to other exoMIPs: the detectability of recorded differences between up-stream codes, and the spread or uncertainty in the down-stream codes.

The objective of this paper is to present the MALBEC experimental protocol that will be used to compare 10 already participating models. A clear and descriptive protocol, separated from the result papers, as well as broad participation is essential for a meaningful model intercomparison project. This paper is structured as follows. In Section \ref{sec:motiv}, we outline the motivation for the MALBEC exoMIP, while Section \ref{sec:considerations} describes the physical and computational considerations that came across during the development of the study, as well as the input/output files. In Section \ref{sec:models} we describe the models currently onboard MALBEC, and the experimental protocol is addressed in Section \ref{sec:proto}. The MALBEC repository, the output formats, standards and archiving are discussed in Section \ref{sec:repo}. The summary is given in Section \ref{sec:summary}.

\section{Motivation for the model intercomparison project} \label{sec:motiv}
Radiative transfer (RT) models are used for a large variety of planetary science applications, from solar system objects \citep{Villanueva2018,Villanueva2022} to exoplanets \citep[and references therein]{Waldmann2015,Arney2017,Caldas2019,Fauchez2019,Lustig-Yaeger2019,Wunderlich2020,Konrad2022}. For exoplanet applications, the number of RT codes has been rapidly rising over the years, following the sharp increase of the number and diversity of planets discovered, alongside the development of various ground-based and space-based observatories either dedicated or significantly oriented toward exoplanet detection and characterization. In this context, it is essential to ensure that RT models are producing similar results to not potentially bias both the data prediction and interpretation.

A detailed model intercomparison study has many benefits both to users and developers of the code and to the broader exoplanetary community. The benefits of such a study are twofold: on the one hand it helps to make an assessment of the differences in models predictions and potentially attach an uncertainty to the simulations, and on the other hand, it can help the different RT codes to be more consistent across the community, highlighting the most important functionalities, asserting what elements would benefit from further development, and overall improving every model fidelity.

By allowing each RT code to work with its built-in functions and specific engine, i.e., without constraining every single parameter nor forcing a particular analysis pipeline (see further sections for precision about the inputs), this intercomparison provides an objective image about each code's specifics and characteristics. Note that this strategy is different from the one adopted by \citet{Barstow2020} where each model had to use exactly the same set-up and only the engines themselves were compared. MALBEC aims at identifying the sources of disagreement among the participant codes, quantifying the differences, and triggering discussions in a constructive environment, towards improving the respective performances. As there are no ground-truths in these types of simulations, understanding differences between codes can provide much insight, and promote the unification of certain schemes and methods that increase RT fidelity. Importantly, the framework developed in MALBEC can be applied by any user who would want to work with one of these models as a benchmarked example, establishing a starting reference point.

\section{Physical and computational considerations}
\label{sec:considerations}
The different radiative transfer models participating in MALBEC inevitably have different strategies to capture the physical phenomena, different emphasis (e.g., direct imaging, transit, gas giants, Earth-like planets), as well as different computational descriptions of the RT phenomena. In the first part of this section, we discuss the physical processes underlying the RT calculations, with a particular focus on the processes where the codes differed, where they had to be aligned, or where the input standards were notably different. Subsequently, we came across a number of computational approaches that were substantially different for some of the participating codes (e.g., atmospheric layering and pressure/temperature definitions), which will be covered in the next section. In some cases, small differences in the assumed constants (i.e., size of the Sun, the gravitational constant) resulted in large differences in transit depths. In order to avoid discrepancies on constants, all simulation parameters defined in the MALBEC files are in absolute standard units.

\subsection{Spectroscopy and Physical Phenomena}
Throughout the wavelength range under investigation in MALBEC, many processes affect the radiation propagating through an atmosphere. Table \ref{tab:phys_eff} highlights these and in which spectral regions they are most relevant. Spectroscopic linelists are at the core of radiative transfer calculations, and in many cases, they are the dominant source of discrepancy between models. Considerations regarding the inclusion of isotopes, of high-energy lines for warm atmospheres, of lineshape models and wing-cutoffs can have substantial effects on the computed spectra. In addition, how these spectroscopic linelists are integrated into the RT model, such as line-by-line, cross-sections and correlated-k methods, can lead to different levels of fidelity.

Based on the spectroscopic parameters (including strength and energies) of the different ro-vibrational molecular transitions, the absorption of a particular gas can be determined for a temperature-pressure combination. The absorption lineshapes also depend on the properties of the dominant collisional partner in the atmosphere (e.g., air, CO$_2$, H$_2$, He). Different databases are available, with HITRAN \citep{GordonEtal2022jqsrtHITRAN2020} being one of the most comprehensive databases, containing 55 molecules and 148 isotopologues at different collisional regimes, and about 300 in experimental cross-sections. The focus of HITRAN is Earth's atmospheric conditions, and for that reason molecular states and transitions that are relevant at different atmospheric temperatures and pressures are not always included. For higher temperatures, the HITRAN team created the HITEMP database \citep{RothmanEtal2010jqsrtHITEMP2010}, which contains spectroscopic information of amongst other \ce{H2O}, \ce{CO2}, CO \citep{Li2015}, \ce{N2O} \citep{Hargreaves2019} and \ce{CH4} \citep{HargreavesEtal2020apjsMethaneHITEMP}, molecules particularly relevant for exoplanet atmospheres. Additionally, the ExoMol repository \citep{Tennyson2016, Chubb2021} is specifically focused on exoplanets, brown dwarfs, and cool stars, and hosts a large spectroscopic databases of molecules, which are particularly suited for high temperature and non-LTE regimes. Specialized databases that combine several repositories also exist, such as HELIOS-K \citep{Grimm_2021} [https://dace.unige.ch/opacity], PSG databases \citet{Villanueva2022} [https://psg.gsfc.nasa.gov/helpatm.php], or MAESTRO \citep{Batalha2020} [https://science.data.nasa.gov/opacities/app].

RT codes have different ways of handling linelists, and line shape models - including different assumptions regarding line wing cutoffs - can have substantial impact on the computed spectra. The most direct method is to calculate the absorption by line-by-line shape modeling, which is very  accurate but computationally expensive, and it is not always possible in the UV/optical since linelists typically do not accurately cover this region. Some RT codes rely on UV/optical cross section tables to complement the line-by-line/correlated-$k$ tables. The MPI-Mainz provides extensive cross section data for many species in this spectral range \citep{keller-rudek_mpi-mainz_2013}. For low resolution studies (i.e., resolving powers less than 5000), the absorbance is often saved in correlated-$k$ tables, which capture the distribution of absorbance in a spectroscopic bin in a selected number of points. Correlated-$k$ tables are pre-calculated for a grid of temperature and pressure combinations and interpolated upon running RT calculations. Alternatively, cross-section data can be used as well, which are pre-calculated absorption spectra for particular temperature-pressure conditions, but the RT calculation need to be performed at very high spectral resolution to ensure high fidelity \citep{garland2019}. Isotopes are available both in HITRAN and ExoMol. Even when using the same databases/methods, the inclusion and treatment of isotopologues among the RT codes varies greatly, mainly due to the assumed linelist, and the considered isotopic abundances. For most isotopes, spectral shifts are minor and are typically lost under the main isotope's absorbance features (e.g., $^{13}$CH$_4$), but in some cases, absorbance of the heavier isotopes is in absorbance minima of the main isotope and can thus be important to consider (e.g., HDO). 

At short wavelengths, Rayleigh Scattering has a strong wavelength dependency, resulting in a smooth increase from the near-IR, visible to UV, following $\lambda^{-4}$. The magnitude depends on the polarizability of the atoms/molecules, but in principle all (dominant) gases in an atmosphere scatter light in this way. The formulation used in RT codes is often the one described in \citet{2005Sneep_Ubachs}. The tabulated values are available from NIST\footnote{https://cccbdb.nist.gov/pollistx.asp}, yet different codes implement this process in slightly different ways. In general we have encountered relatively small differences on this process across the ``casks" compared here. 

Primarily in the infrared, but also in the UV and in the visible, Collision Induced Absorption (CIA) arises from temporary bi-molecular complexes, resulting in relatively broad features whose intensities scale quadratically with pressure. Description of the phenomena and how to calculate the corresponding absorption cross-sections can be found in \citet{2019Karman}. The HITRAN CIA repository \footnote{https://hitran.org/cia/} lists the most generally used CIA lists. Depending on the main emphasis of each RT model, the number of CIA databases and considered repositories varies greatly. RT models with an emphasis on Earth-like planets implement CIA lists well beyond HITRAN, including water continuum and capturing several \ce{N2}-x and \ce{O2}-X CIA bands, while gas-giant RT models mainly consider several theoretical and lab-based CIA databases for \ce{H2}-X and He-X pairs. 

\begin{table}
\centering
\caption{List of main processes impacting the simulated spectra of planets and investigated by this investigation.}
\begin{tabular}{l l l } 
\hline
\hline
Spectroscopic               & Affected Region         & Notes \\
\hline
Rayleigh Scattering         & $<$1.5 {\textmu}m         & Different implementations exists \\
Molecular absorption:    & All wavelengths   & \\
 • Isotopes & Specific features and deine molar masses & \textit{e.g.} HDO, \ce{CO2} isotopologues\\
 • Line-by-line & All wavelengths & Main sources, HITRAN, HITEMP, ExoMol \\
 • Correlated-$k$ & All wavelengths & Each model implements these differently \\
 • Cross-sections & Critical at UV/Optical & Complement linelists with cross-sections \\
 
Collision Induced Absorption & Continuous & Main sources: HITRAN, MT\_CKD, theory \\
\hline
\hline
\label{tab:phys_eff}
\end{tabular}
\end{table}

\subsection{Computational considerations, definitions of terminologies, and input/output specifications}
During the development of this protocol, we investigated a number of different treatments of the computational aspects of the radiative transfer calculations, to understand the fundamental differences between codes. To ensure all models interpreted the different tests (``casks") similarly, a common input file was agreed upon. This “configuration” file, called \textit{the MALBEC file}, contains all the information required to calculate the different test cases, in a consistent format across the different ``casks", and can be ingested by each RT code within MALBEC.

Not all radiative transfer codes discretize the atmosphere in the same way, so a method was developed to homogenize the layering schemes (pressure and temperature versus height) for all RT codes. In addition to providing anchor points at specific altitudes to describe the local atmospheric pressure, temperature, and molecular/aerosols abundances, we also provide polynomial fits to these quantities, so each code can discretize the atmosphere in any preferred manner. This is relevant both for the radiative transfer calculations in each layer, and for the path lengths of light through each layer (i.e., ray tracing). Particularly for ray-tracing calculations in spherical refractive atmospheres, it is important to sample at sufficiently high resolutions, either by having enough layers or by creating sub-layers. If insufficient layers are used this may lead to large differences. In the configuration file we provide “anchor” values, defining the specific pressure, temperature and abundances at specific points in the atmosphere. These points quantify the properties of the atmosphere at that exact pressure point, yet when performing ray-tracing calculations to determine the path length of individual layers, each model may sub-divide the light path between anchor points into much finer points by using a sub-layering scheme in order to ensure an accurate sampling of the large variability of the parameters in-between “anchor” points. Additionally, we also provide analytical fits (3rd order vs. log$_{10}$ pressure [bar]) to those profiles, so the user can create their own vertical discretization, ensuring accurate results with their respective model. 

The configuration file defines the following:
\begin{itemize}
\item \textbf{Planetary and Stellar parameters:} Planetary radius [m], surface gravity [m/s$^{2}$], phase at which the planet is observed, planet temperature [K], albedo, emissivity, star-planet distance [m], stellar type, stellar temperature [K] and stellar radius [m]. Albedo and emissivity are defined to be constant across the simulation wavelength range, and the surface scattering model is assumed to follow the Lambert diffuse principle.
\item \textbf{Opacity sources opacity sources used:}  linelists, CIAs, Rayleigh, etc. 
\item \textbf{Atmosphere-columns}: this entry specifies the variables included in the simulation and the anchor points. Pressures are in [bar], temperatures in [K], mean molecular weight in [g/mol], height in [km], and molecular abundances in [mol/mol]. 
\item \textbf{Layers}: the specifics of each layer, containing the pressure, temperature, mean molecular weight, altitude, and molecular abundances following the definitions described above.
\item \textbf{Analytical functions layering:} For a number of relevant species, 3rd order polynomial parameters describing the abundance with the log$_{10}$ of the pressure are given.
\item \textbf{Bottom of the atmosphere (BOA):} the atmosphere is defined to start at the first anchor point, with surface properties as listed in the configuration file (surface temperature, albedo, emissivity). The planetary gravity and radius refer to this point.
\item \textbf{Top of the atmosphere (TOA):} the atmosphere is defined to end at the last anchor point, and therefore the atmospheric radiative transfer is only computed between BOA and TOA.
\item \textbf{Spectral range and sampling:} four parameters define the range and resolution of each simulation. The minimum wavelength defines the center wavelength of the first “pixel”, while maximum wavelength defines that of the last “pixel”. The separation between pixels can be a fixed wavelength if the unit is “{\textmu}m”, while if the unit is ``RP”, the separation assumes a constant resolving power ($\frac{\lambda} {\delta\lambda}$). 
\item \textbf{Output units:} for transit simulations, the output model units are in ($\frac{R_{planet}}{R_{star}})^2$, while for direct imaging simulations, the model outputs are in [W/sr/m$^2$/{\textmu}m].
\item \textbf{Opacities:} the models should employ all opacity terms suitable for each test case, for instance high-energy linelists would be beneficial when simulating hot atmospheres as in case T3B (see below). This includes molecular opacity (for all isotopes), Rayleigh scattering/extinction, water continuum (e.g., MT$\_$CKD) and Collision-Induced-Absorptions (CIA). Molecular opacities should be generated employing the most appropriate linelist available to each model for that collisional regime (e.g., \ce{H2}, air broadening). All simulations as part of this intercomparison are at moderate resolutions (with resolving powers equal or lower than 500), and therefore correlated-$k$ or other pre-computing methods are encouraged.
\end{itemize}

\section{Description of the models participating in MALBEC}\label{sec:models}
The MALBEC intercomparison is open to the whole community, with the framework of this initiative being defined among a group of diverse radiative-transfer teams and models. The RT models are quite different in their scope and functionality, which is important when establishing and comparing results across a broad range of regimes, as required by the exoplanetary community. Importantly, not all models are required to address all intercomparison experiments, and in this section we list the capabilities and characteristics of each of these initial RT models.

\begin{table}
\centering
\caption{Radiative transfer models participating in MALBEC, along with their point-of-contact (POC) and notable model references.}
\begin{tabular}{c c c }
\hline
\hline
Code & POC & References \\
\hline
APOLLO & Alex Howe & \cite{Howe2022} \\
BART & Michael D. Himes & \cite{HarringtonEtal2022psjBART1, CubillosEtal2022psjBART2} \\
& & \cite{BlecicEtal2022psjBART3} \\
gCMCRT & Elspeth K.H. Lee & \cite{Lee2019, Lee2022}\\
petitRADTRANS & Eleonora Alei & \cite{Molliere2019} \\
PSG/PUMAS & Geronimo L. Villanueva & \cite{Villanueva2018, Villanueva2022}\\
Pytmosph3r & Jeremy Leconte & \cite{Caldas2019} \\
SMART & Michaela Leung, Samantha Gilbert, & \cite{MeadowsCrisp1996}\\
 & Gabrielle Suissa, Edward Schwieterman & \\
SOCRATES & Mei Ting Mak, James Manners,& \cite{Lines18_exonephology}\\ 
& Denis Sergeev, Nathan Mayne & \\
TauREx 3 & Estelle Janin & \cite{Waldmann2015a, Waldmann2015b, Al-Refaie2021A3.1}  \\
\texttt{PICASO} & Thea Kozakis & \cite{Batalha2019,Batalha2021} \\
\hline
\hline
\label{tab:models}
\end{tabular}
\end{table}

\subsection{APOLLO}
APOLLO is a 1D radiative transfer and MCMC spectroscopic retrieval code designed for modularity and internal model intercomparisons. The main description of APOLLO can be found in \cite{Howe2022}, and the model can be used to synthesize both transit and emission spectra. It is built on a philosophy of free retrieval that is agnostic about specific microphysics and as such allows the user to select from several options for prescriptions for temperature-pressure profiles, cloud structures, and molecular inventories. While multiple radiative transfer prescriptions are also available, by default, APOLLO simulates radiative transfer (reflected and emission) through an atmosphere using the two-stream approximation \citep{Toon1989} with hemispheric closure \citep{Meador_Weaver_1980}. APOLLO uses the \cite{Freedman2014} cross section tables on a species-by-species basis. A weighted sum of the cross sections for each species is then computed based on the mixing ratios to compute optical depths layer by layer, including CIAs and aerosols. For transit spectroscopy, the starlight is assumed to be parallel. For emission spectra, the emitted light is integrated over an eight-point Gaussian quadrature. APOLLO can perform retrievals by an MCMC method using the emcee Python package \citep{Foreman-Mackey2013}. APOLLO also includes an ``ensemble'' method to rapidly generate and evaluate regularly-spaced forward model grids.

\subsection{BART}
\label{sec:BART}
BART \citep{HarringtonEtal2022psjBART1, CubillosEtal2022psjBART2, BlecicEtal2022psjBART3} is an open-source exoplanet retrieval code which pairs the {\tt transit} RT code \citep{Rojo2006PhD} with the Multi-Core Markov-Chain Monte Carlo  \citep[MC3;][]{CubillosEtal2017apjRednoise} Bayesian framework. BART performs 1D radiative transfer under the assumption of hydrostatic balance and LTE. For transmission spectra it performs ray tracing, and for emission spectra it assumes a plane-parallel atmosphere. Its programmatic design focuses on flexibility, enabling BART to replicate the setups of other codes and achieve agreement in results \citep[e.g.,][]{HimesHarrington2022apjWASP12b}. The RT routines are implemented in C for speed, while BART is implemented in Python for ease of use and portability. Both {\tt transit} and BART are generally executed via configuration files, where the user may specify the various parameters of the run, including the spectral range and resolution, atmospheric species and profiles, Rayleigh extinction and/or cloud models, and the observing geometry (transit, eclipse, or direct observations). At present, BART does not consider multiple scattering when computing emission or transmission spectra. {\tt Transit}'s package to ingest linelists allows for HITRAN/HITEMP \citep{RothmanEtal2010jqsrtHITEMP2010, HargreavesEtal2020apjsMethaneHITEMP, GordonEtal2022jqsrtHITRAN2020}, ExoMol (\citealp{Tennyson2016}; \citealp{Chubb2021}; via Repack, \citealp{Cubillos2017apjRepack}), TiO \citep{Schwenke1998fadiLineListTiO}, and the \ce{H2O} list of \citet{PartridgeSchwenke1997jcpLineListH2O}. {\tt Transit} utilizes a line-by-line opacity sampling approach to reduce the sizes of extensive linelists while still retaining an accurate RT calculation (see \citealp{Rojo2006PhD}, \citealp{HarringtonEtal2022psjBART1}, and \citealp{CubillosEtal2022psjBART2} for more details on other computational cost-saving measures).  In Figure \ref{Fig:T3B} of test cask T3B, we utilize the HITEMP linelists for H$_2$O and CH$_4$; HITRAN CIA databases for H$_2$-He, H$_2$O-H$_2$O, CH$_4$-CH$_4$, CH$_4$-He, and H$_2$-CH$_4$; and the H$_2$-H$_2$ CIA database of \citet{Borysow2001jqsrtH2H2CIA, Borysow2002aapH2H2CIA}.During a retrieval, BART calculates high-resolution spectra and bins them according to the instruments used to observe the exoplanet.  MC3 calculates the effective sample size of the posterior resulting from the inference as well as its induced uncertainty in credible regions determined from that posterior (see \citealp{HarringtonEtal2022psjBART1} for more details).

\subsection{gCMCRT}
gCMCRT is an open source 3D spherical geometry hybrid ray tracing and Monte Carlo radiative-transfer code which uses GPUs to accelerate calculations. The main descriptions of gCMCRT can be found in \citet{Lee2019} and \citet{Lee2022}. gCMCRT is primarily aimed at accurately post-processing 3D GCM data, especially when multiple scattering from cloud particles is important. gCMCRT directly simulates the path of photon `packets' (usually $\sim$ 10$^{6}$) through the prescribed 3D spherical geometry structure. This is a sub-grid model, with packets traversing through the 3D volume of each cell.
For transmission spectroscopy, packets are fired at the dayside of the planet and traced through the transmission limbs, which are then used to estimate the transmission spectra integral through random sampling. For emission spectra, packets originate in the 3D spherical volume itself, and are traced towards a certain viewing angle (planetary phase). The fraction of energy escaping towards this direction is then tracked and used to calculate the emission spectra. For albedo calculations, the scattered fraction of the incoming starlight is calculated towards a certain viewing direction. gCMCRT can also perform multiple-scattering in 3D geometry by sampling scattering phase functions (for example, the Rayleigh or Henyey-Greenstein phase functions) when a packet undergoes a scattering event. gCMCRT contains a well tested opacity mixer (optools), which can mix and prepare gas phase abundances, correlated-$k$ tables, line-by-line, CIA, Rayleigh and cloud opacities and scattering properties. gCMCRT is able to perform high-resolution spectral modelling including Doppler shifting of lines, allowing simulated GCM data to be used as part of high-resolution cross-correlation studies.

\subsection{petitRADTRANS}
petitRADTRANS (pRT for short) is a publicly available package to calculate transmission and emission spectra of clear and cloudy exoplanets \citep[see][for more details]{Molliere2019,Molliere2020}. The core routines are written in FORTRAN and wrapped in a Python package. The model can run at high-resolution ($\lambda/\Delta\lambda\le10^6$) using a line-by-line treatment, as well as at low resolution ($\lambda/\Delta\lambda\le10^3$) using a correlated-$k$ approach. Opacities can be rebinned at any resolution needed by the user, either directly in petitRADTRANS (for high resolution), or through the \texttt{exo-k} package \citep{2021A&A...645A..20L} which is called internally (for low resolution). The default set of opacities provided by the maintainers includes opacities from HITRAN/HITEMP and ExoMol, generally spanning a temperature range between 80 and 3000 K, and a pressure range between $10^{-6}$ and $10^{3}$ bar. Further opacity tables can be downloaded from ExoMol directly in a pRT-friendly format, or calculated by the user through the ancillary scripts provided.

petitRADTRANS allows the user to consider different cloud treatments, such as gray clouds, wavelength-dependent power laws, or using opacity (in $cm^2/g$) of real condensates of various compositions and shapes. Arbitrary opacity functions and particle setups can also be employed. Users can also provide low-res (pseudo-continuum) opacities to petitRADTRANS to be included in the calculation, thus customizing the cloud treatment. petitRADTRANS includes the calculation of Rayleigh scattering and collision-induced absorption. It was recently updated to include the surface direct and thermal scattering for rocky exoplanets \citep[see Appendix A in][for details]{Alei2022}. Specifically for the computation of multiple scattering of reflected starlight and thermal emission, pRT uses the Feautrier method, which is a third-order method that allows the treatment of the radiative transfer equation in the diffusive regime, and for which pRT iterates to solve for the multiple scattering source by employing the ALI and Ng acceleration. Users can specify wavelength-dependent surface reflectivity and emissivity, allowing the scattering treatment of various surface compositions of rocky exoplanets. In its latest version, it also allows the user to run retrievals using the nested sampling approach. Thanks to the MALBEC collaboration, opacities of additional isotopes for many atmospheric absorbers (e.g., \ce{CO2}, CO, \ce{H2O}) were computed, as well as the collision-induced absorption tables for water (MT\_CKD, \citealt{kofman_absorption_2021,2012Mlawer}). These will be soon included in the main opacity repository of petitRADTRANS. For the purpose of this intercomparison, the version of petitRADTRANS was frozen to 2.6, though the package is in the process of being upgraded to version 3, with additional features and quality-of-life improvements.

\subsection{Planetary Spectrum Generator (PSG)}
The Planetary Spectrum Generator, \citet{Villanueva2018, Villanueva2022}, is a publicly available radiative-transfer suite, generalized to any type of planetary atmosphere/surface across a broad range of wavelengths (50 nm to 100 mm, UV/Vis/near-IR/IR/far-IR/radio) and observational geometry (e.g., transit, direct-imaging, in-situ, orbiter, limb). The suite includes a 3D orbital calculator for most bodies in the Solar system and all confirmed exoplanets, and also includes a broad range of atmospheric and chemical models. The ray-tracing integration across the atmosphere is performed in PSG via a sub-layering spherical/refractive iterative algorithm, enabling high-fidelity results even when operating with a limited number of layers. Several radiative-transfer modules operate within PSG, including CEM (Cometary Emission Model), a non-LTE exospheres solver, and PSGDORT, the multiple-scattering radiative-transfer module of PSG that employs the discrete ordinate method to model atmosphere/surface scattering processes. PSGDORT is based on DISORT v2.1 \citep{stamnes2000} which were adapted for non-LTE and optimized to operate with a variety of spectral grids (e.g., line-by-line, correlated-k, surface scattering grids) as employed by the PSG radiative transfer algorithm. PSGDORT also includes correction for pseudo-spherical geometry as described by \citep{dahlback1991}. The PSG radiative transfer modules take into account Rayleigh and Raman scattering, refraction, non-LTE processes, molecular/atomic absorptions, and collision induced absorption processes. Ro-vibrational absorption are calculated line-by-line for resolving powers higher than 5000 and correlated-$k$ tables are used for lower resolutions.

PSG operates with several databases, including the latest HITRAN/HITEMP distribution (linelist, CIAs, cross-sections), GEISA, and ExoMol. For the PSG simulation results presented in Figure \ref{Fig:T3B} of cask T3B, we utilized correlated-k tables based on the HITEMP linelists of H$_2$O and CH$_4$. PSG complements the UV/optical range, by integrating UV/optical cross sections for dozens of species collected from a range of spectral databases. Most of the UV cross-sections originate from the MPI-Mainz Spectral Atlas \citep{keller-rudek_mpi-mainz_2013}, which have been parsed, combined and formatted to provide a comprehensive coverage of the 0.01 to 1 {\textmu}m wavelength range. Additional cross-sections include those of \ce{O3} by \cite{serdyuchenko2014}, \ce{CO2} by \cite{venot2018}, the Herzberg \ce{O2} continuum bands as well as the \ce{O2}–\ce{O2} absorption bands (Wulf bands) by \cite{fally2000}, the Herzberg \ce{O2} band system \citep{merienne2001}, and several \ce{O2}-X CIAs \citep{fauchez2020}. The MT\_CKD water continuum is parameterized in PSG by computing CIAs tables for \ce{H2O}-\ce{H2O} and \ce{H2O}-\ce{N2} from that water model \citep{kofman_absorption_2021,2012Mlawer}. PSG also includes a retrieval framework, optimal estimation and Multinest methods, and a noise simulator. The tool can be operated in several ways: via a public web interface\footnote{https://psg.gsfc.nasa.gov}, via a flexible application program interface (API) accessible from any scripting language (e.g., python, IDL), and it can be also operated by an installable suite via the Docker virtualization system. Additionally all opacity tables, spectroscopic databases, and core fundamental source codes are publicly availble at the site and the PSG github respository.

\subsection{\texttt{Exo\_k / Pytmosph3r}}
Originally developed to easily handle, convert, and interpolate radiative opacities from various sources, the \texttt{Exo\_k} python library \citep{2021A&A...645A..20L} now includes a 1D radiative transfer module able to model the transmission and reflection/emission spectra of a wide range of atmospheres. Thanks to its original philosophy, \texttt{Exo\_k} can include any molecule or opacity source for which opacities can be provided in the form of either cross sections or correlated-$k$ coefficients tables in one of the many formats handled by the library. In particular, data can be provided at any resolution as the library implements a novel algorithm able to bin-down even $k$-coefficient tables \citep{2021A&A...645A..20L}. A high numerical efficiency is ensured by the use of the the ``just in time" compilation capability of the \texttt{numba} library.

The reflection/emission algorithm is an improved version of the two-stream framework from \citet{TMA89} that accounts for multiple scattering and scattering asymmetry (see the appendix of \citet{CTB22} for details). Mie scattering can be modeled for any type of aerosols by providing the aerosol optical properties. The change in atmospheric scale height due to both gravity and mean molecular weight variations are taken into account, both in emission and transmission.

\texttt{Pytmosph3r} is the 3D implementation of \texttt{Exo\_k} that can be used to model observables from both parameterized 3D atmospheric structures and outputs from a 3D global climate model \citep{Caldas2019, FZP22}. Using \texttt{Exo\_k} for all the opacity calculations, it inherits all the aforementioned features. In transmission, rays of light are shot through the limbs of the planet, possibly crossing many atmospheric columns of the atmospheric model and thus accounting for its heterogeneity, both along and across the limb. In emission/reflection, the two-stream algorithm from the 1D model is used in each atmospheric column to compute the radiative intensity in the direction of the observer -- thus accounting for limb darkening. Intensities are then weighted by solid angle and integrated over the visible sphere to yield the observed flux at the desired observer's location. 

\subsection{SMART}
The Spectral Mapping and Radiative Transfer (SMART) code is a highly flexible one-dimensional, plane-parallel, line-by-line, multi-stream, and multiple-scattering radiative transfer model originally presented by \cite{MeadowsCrisp1996} and \cite{Crisp1997}. SMART solves the radiative transfer equation using the discrete ordinates method through the DISORT FORTRAN code \citep{Stamnes1988}. SMART ingests gaseous absorption coefficients generated by its Line-By-Line Absorption Coefficients (LBLABC) companion model, as well as optional supplemental cross sections typically used to compensate for gaps in line-list opacity data, and separately sourced collisionally-induced absorption (CIA) coefficients. LBLABC uses the HITRAN linelists to calculate absorption coefficients \citep{GordonEtal2022jqsrtHITRAN2020}. Usually cross-section data derived from the MPI-Mainz database \citep{keller-rudek_mpi-mainz_2013} or other sources is used for UV-Vis wavelengths where linelist data is often absent; however, cross-section data can also be used at any wavelength where appropriate. The same code is used for both shortwave and longwave radiation. The version of SMART employed here uses the ray tracing model of \cite{Robinson2017} to simulate transit transmission observations, which includes the effects of refraction. SMART has been repeatedly verified against terrestrial solar system objects \citep[e.g.][]{Tinetti2005,Robinson2011, Arney2014} and has a long history of being used to generate exoplanet spectra \citep[e.g.][]{Charnay2015, Lincowski2018,Lustig-Yaeger2019}. SMART is highly customizable with user selected input gases and high internal resolution. The model can also be configured for clouds and aerosols. 

\subsection{SOCRATES}
``Suite Of Community RAdiative Transfer codes based on \cite{Edwards1996}'' (SOCRATES) is an open-source radiative transfer code with a two-stream multiple scattering solver for reflected starlight and thermal emission. It uses the correlated-$k$ method to solve for gaseous absorption, with further details covered in \citet{Edwards1996} and \citet{Manners22}. Opacity data are sourced from a range of databases such as HITRAN, HITEMP and ExoMol \citep[see][for full details]{Amundsen2014,Amundsen2016,Goyal2018}. For cask T3B, correlated tables were constructed from ExoMol linelists. At the moment, the SOCRATES code was written to calculate transmission spectra for the global grid from the Unified Model (UM) --- a 3D GCM developed and used by the Met Office, and at present SOCRATES cannot calculate transmission spectra outside a GCM. However, a standalone version of this calculation is planned for a near-future release of SOCRATES. Within each GCM column the transmitted solar flux is calculated using a spherical shell geometry, replacing the standard plane-parallel assumption \citep[detailed description can be found in][]{Lines2018}. The total optical depth through the atmosphere is found by summing the mass along the slant path through each shell multiplied by the corresponding mass extinction coefficient for the layer. Each model column is treated independently and each layer is assumed to be homogeneous with identical optical properties in that column. The fluxes are calculated in the column facing away from the star in the limb. The resulting fluxes in each column are added up to represent the full transmission spectrum. With this approach, the variations of fluxes in the limb perpendicular to the observer can be fully captured but those along the line-of-sight can only be approximated. The assumption of homogeneous spherical shell might also lead to a bias if there are variations across the terminator.

\subsection{TauREx 3}

TauREx (Tau Retrieval for Exoplanets) is an open-source program initially developed by \citet{Waldmann2015a,Waldmann2015b} for both transmission and thermal emission spectroscopy. The current version in use -- now completely Python-stacked -- corresponds to TauREx 3 and is substantially described in \citet{Al-Refaie2021A3.1}. TauREx is a fully Bayesian, line-by-line radiative transfer and retrieval framework. It can be used with any molecular line-lists in hdf5 and in pickle format, including the latest data from the Exomol project \citep{Tennyson2016, Chubb2021}, HITEMP \citep{Rothman2014} and HITRAN \citep{Gordon2017}. Its modular, object-oriented architecture, combined with an extensive TauREx 3 Python library, makes it particularly flexible and customisable to the user. TauREx models a 1D-atmosphere and computes the path integral from a single cross-section where each (interpolated and weighted) opacity has been fused. The opacity integral is computed based on a finite difference scheme. In addition, TauREx provides a full implementation of Rayleigh and Mie scattering via extinction, as well as simple (Grey) cloud models. It has been particularly tailored for the modeling and interpretation of transit spectra, so it only considers extinction processes and it does not solve for scattering emission when computing thermal emission. The user builds both transmission and emission models by defining a temperature profile, pressure parameters, a chemistry model, and contributions to the optical depth (molecular absorption, CIA, Rayleigh, Mie scattering, simple clouds). The wavenumber grid of the forward model is selected at runtime from whichever loaded opacity with the highest resolution. Every other opacity is then resampled to the chosen opacity's grid before any computation begins; this grid becomes the native grid of the forward model. 

In this study, the T-P profile, as well as chemical abundances at each layer, were provided as inputs (thus not requiring the use of any TauREx chemical model). For the TauREx simulation results presented in Figure \ref{Fig:T3B} of case T3B, we used ExoMol absorption cross-sections for H$_2$O \citep{polyanski} and CH$_4$ \citep{Yurchenko2017}, CIA opacities for H$_2$-H$_2$ \citep{Fletcher2018}, H$_2$-He \citep{Abel2011}, CH$_4$-X (from HITRAN), as well as Rayleigh scattering contributions \citep{Cox2015}.

\subsection{\texttt{PICASO}}

\texttt{PICASO}, the Planetary Intensity Code for Atmospheric Scattering Observations, is a radiative transfer code originally designed for reflected light planetary spectra \citep{Batalha2019}, and has since been expanded to include calculations of transmission and emission planetary spectra \citep{Batalha2020,Batalha2021}, 3D spectral modeling \citep{Adams2022}, phase curves \citep{Robbins-Blanch2022}, 1D climate modeling \citep{Mukherjee2023}, as well as fitting spectroscopic data to models \citep{JWSTers}. The code is written in Python and publicly available on GitHub\footnote{https://github.com/natashabatalha/picaso}, and has been used in a wide range of studies for exoplanets and brown dwarfs (e.g., \citealt{Mukherjee2021,Foote2022,Lew2022,Kozakis2022}). \texttt{PICASO} is based off codes originally described in \cite{McKay1989}, \cite{Marley1999}, and \cite{Cahoy2010}, being built off of methods from \cite{Toon1977,Toon1989} to compute reflected starlight and thermal emission as observed at full phase, with updates from \cite{Cahoy2010} for observations at any phase angle and including multiple-scattering via the two-streams approximation. 

The code is accompanied by a database of opacities resampled down to R = 20,000 and 60,000 from line-by-line calculations on the order of R$\sim$10$^6$ (see methodology in \citealt{GaribNezhad2021} for details). This database spans wavelengths from 0.3 to 14 {\textmu}m and can either be downloaded in entirety or queried using \texttt{sqlite3}. The resampled opacity database draws from multiple databases such as HITRAN and  ExoMol, along with many other sources for linelists, CIAs, and cross sections, with the code providing helpful tools to extract the relevant references for individual model runs. If desired, \texttt{PICASO} additionally provides functions for creation of correlated-$k$ tables for individual gases. Within the code there are options to add clouds in the form of grey cloud coverage or cloud model output \citep{Rooney2022}, as well as user supplied parameteriz]d arbitrary Rayleigh scattering. 

\section{The MALBEC protocol and experiments} 
\label{sec:proto}

The MALBEC model intercomparison contains three set of experiments: 1) core, 2) transit, 3) direct-imaging experiments, with a total of 12 cases (Table \ref{tab:cases}). All simulations are to be carried across a wide spectral range (0.2 to 20 {\textmu}m) at moderate resolutions (RP=200), and each model is asked to provide results only in regimes and spectral regions for which the models have been previously operated. 

\begin{figure}[ht]
\centering
\resizebox{18cm}{!}{\includegraphics{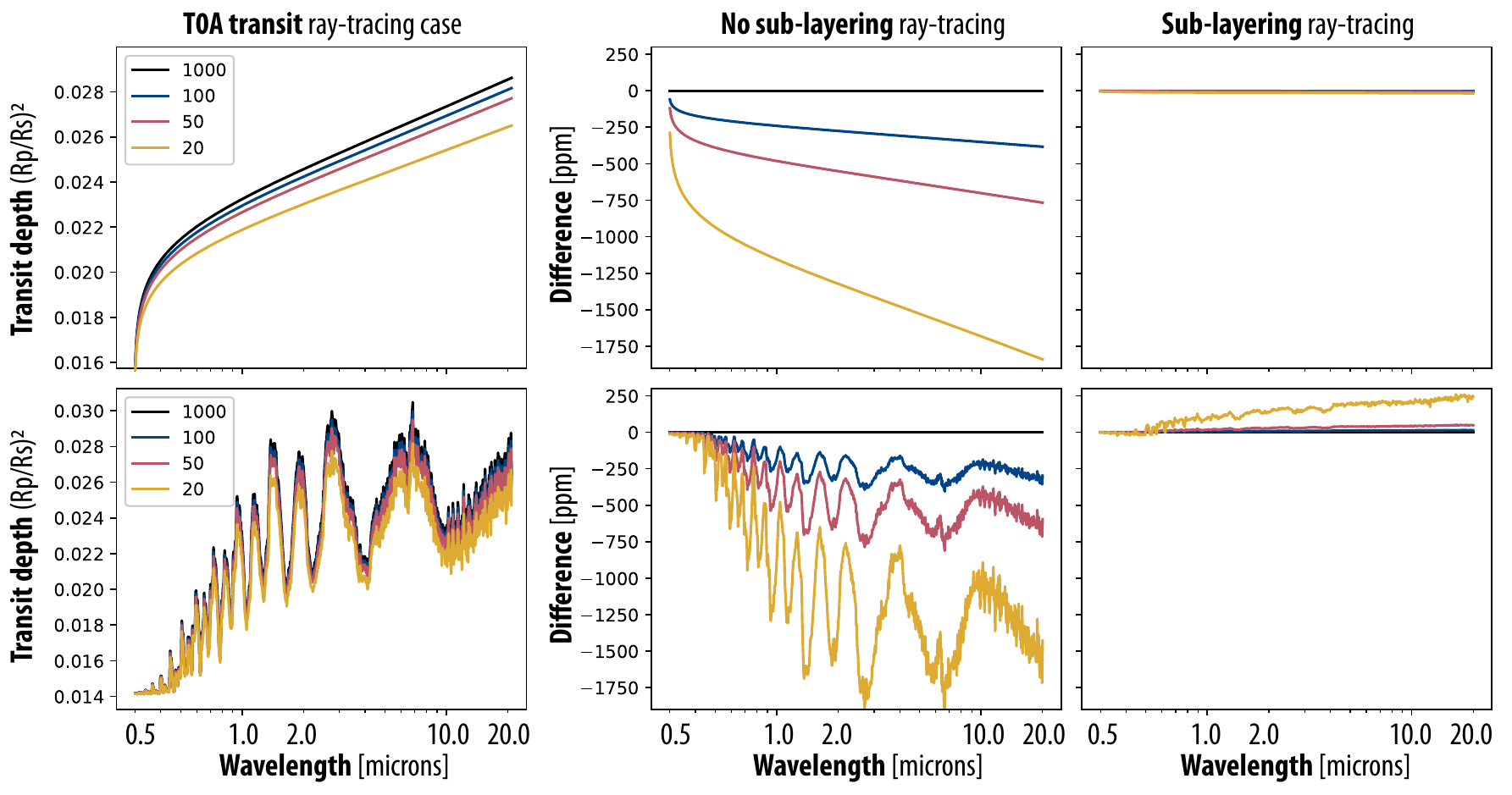}}
\caption{Comparison of the effects of atmospheric discretization and ray tracing on the PSG computation of transit spectra as part of the T0B test cask. The left panels show transit for an hypothetical extended atmosphere assuming a parameterized opacity function (top) and for realistic water opacities (bottom). The different traces were computed  considering different number of layers (20, 50, 100, 1000) for the 100 bar to 1 nbar range. The middle panel shows differences with respect to the 1000 layers solution, which for 50 layers the differences can reach up to 500 ppm at 2 $\mu m$. On the other hand, if the RT model implements a sub-layering algorithm in the ray-tracing scheme these discretization effects can be largely mitigated (right panels).}
\label{Fig:T0B}
\end{figure}

\subsection{The core experiments}

The core experiments (T0A, T3B) are defined specifically to test basic assumptions within each model and to further assist with the homogenization of the results. The purpose of the MALBEC exoMIP is not for all models to agree, but to understand how certain aspects in the parameterization of the models impact the simulations. As such, issues with layering, opacities, refraction, line-lists, scattering considerations, impact the results. As long as each model is parameterized as close as possible to the MALBEC configuration file, then each model result will reveal how parameterizations and modeling issues impact the spectra. 

Some of the most common differences in RT models is how each code discretizes and defines the atmospheric layering. For instance, some models expect values describing the average properties across the layer, while others expect the properties at the border regions. If the number of layers is finite and reduced, then differences in these assumptions can lead to substantial errors. Such type of error primarily originates from the path length the light traces through the atmosphere. To investigate this, the T0A experiment explores how a transit is modelled for the same atmosphere but when considering different numbers of layers (20, 50, 100, 1000). It is demonstrated that if the RT model implements some sort of sub-layering algorithm (e.g., internally dividing the input layers to better capture the change of densities and opacities), these effects can be largely mitigated and practically removed as we can see in Fig. \ref{Fig:T0B} right panels where the transit depth difference between the 1000 layer cases and the lower layer cases tends towards 0 ppm when sub-layering ray-tracing is performed. As noted by \cite{Malik_2017}, sub-layering can also have a strong effect on the numerical convergence of Radiative Equilibrium (RE) and Radiative Convective Equilibrium (RCE) models, which are used to establish and model the thermal structure of planetary atmospheres. 

For the MALBEC intercomparison, the configuration files provide anchor values, values at the border of the layers, and analytical fits to each parameter across the atmosphere, so the modelers can explore and optimize the calculations of the layering as needed. For instance, if the models require layer quantities (not border values), providing straight average values of the two border values would be then more accurate than providing border values.

The T0A is a highly simplified atmosphere designed to test the treatments of layering and sub-layering of the atmosphere, and investigate the effects this has on the final absorbance spectrum.  It is based on the atmosphere of GJ1214b, where the temperature decreases linearly from 1000K at 100 bar to 400K at the top of the atmosphere at 1 nano-bar. The experiment explores two scenarios: a case with a simple parameterized water cross-section, increasing with wavelength, and a case employing realistic water opacities. The simple cross-section increases exponentially from $10^{-26}$ cm$^{2}$/molecule at 0.3 $\mu$m to 8 $10^{-19}$ cm$^{2}$/molecule at 20 $\mu$m, and is designed to probe different altitudes with wavelength and to highlight the layering effects in the final transmission spectrum. More details about the cross-sections employed in this test can be found at the MALBEC repository. In Figure \ref{Fig:T0B}, we show the comparison between the parameterized water absorption (top panels) and using the water cross-section in the bottom panels (\ce{H2O}; No CIA, no continuum, no Rayleigh for simplicity). This serves to show that our simple opacity covers a dynamic range similar to water.

The effects of not including sub-layering in the RT models can be quite noticeable, as shown in the middle panels of Figure \ref{Fig:T0B}, reaching absolute errors beyond 500 ppm at 2 $\mu m$ on the transit radius when no sub-layering is included in the model. Interestingly, the error is not an offset, but more of a trend: the higher in the atmosphere the transit probes, the larger the error. This manifests in smaller errors at weaker water absorption (windows) versus where absorption is stronger (bands).  This discretization and ray-tracing error can be largely mitigated when employing a sub-layering algorithm, as shown in the right panels of Figure \ref{Fig:T0B} and computed with PSG, which subdivides each layer in 10 sub-layers when computing the ray-traced column densities.

\begin{figure}[ht]
\centering
\resizebox{18cm}{!}{\includegraphics{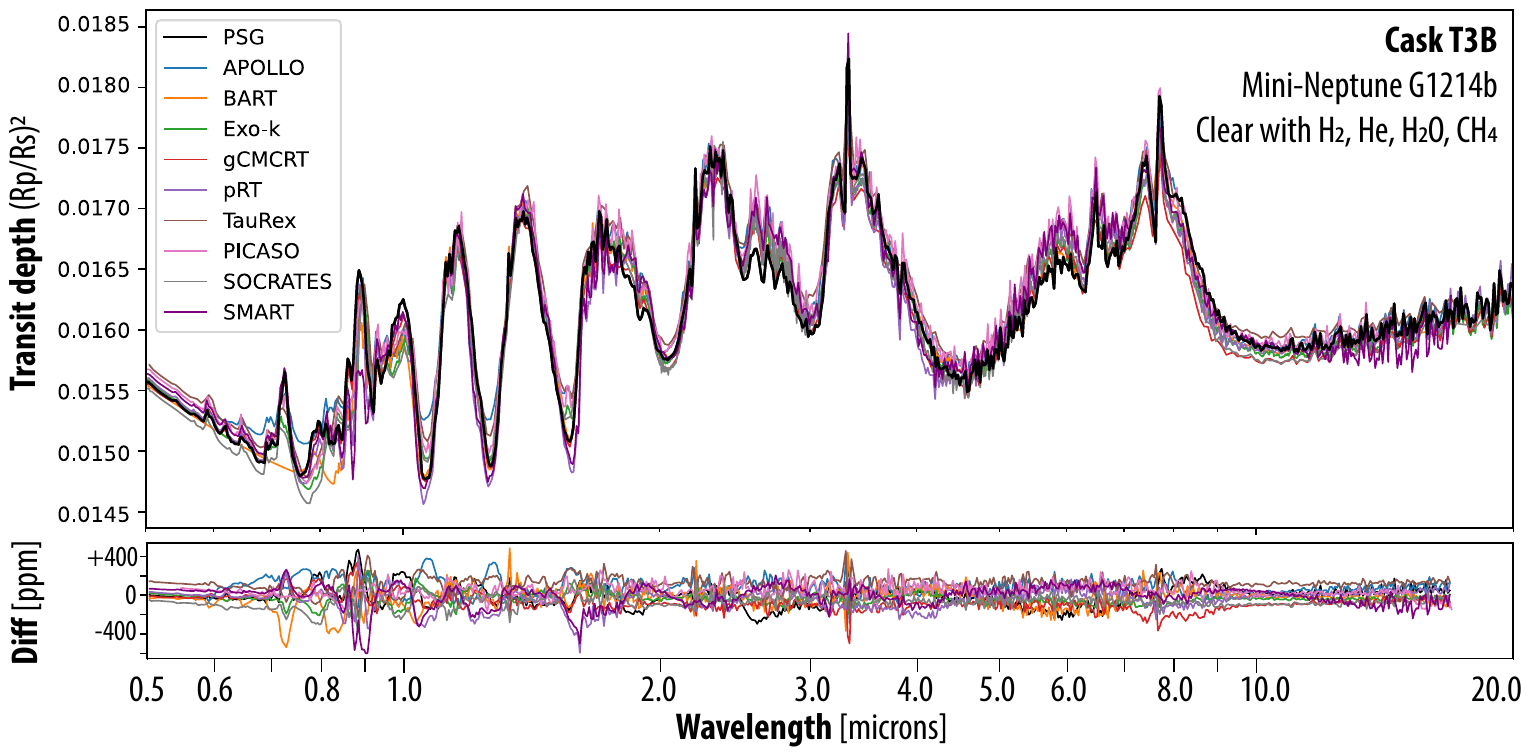}}
\caption{Comparison of preliminary simulation results for test cask T3B. The RT models were run using their typical and core functionalities (e.g., linelists, ray tracing, CIAs) with minimum adaption for this experiment, and were configured to take into account the MALBEC input parameters as close as possible. There is general agreement among the models, with the valleys/peaks in most cases being reproduced by all the models. However, there are specific features in the UV/Optical and at certain wavelengths where the models disagree, which is probably related to the employed linelists and the available CIAs for each RT model. Further revised results with improved parameterizations for each RT model will be reported in a subsequent investigation. See detailed comparisons of the near-infrared region and the optical region in Figures \ref{Fig:T3B_LBL} and \ref{Fig:T3B_UV} respectively. The differences between the models as shown in the lower panel are relative to the mean of all the presented spectra. }
\label{Fig:T3B}
\end{figure}

T3B is a more advanced core experiment, although still with a limited amount of absorbers, but integrating a complex temperature profile, varying molecular abundances and all possible opacity effects (e.g., Rayleigh scattering, CIA, molecular absorptions). In Figure \ref{Fig:T3B}, we show preliminary results for this case computed using the models reported in Table \ref{tab:models} and in section \ref{sec:models}, in which we consider a base pressure of 1 bar, a planetary diameter of 36318 km, and a stellar diameter of 306108 km (more simulation details available at the MALBEC CKAN repository). One can see a general agreement yet notable differences at certain spectral regions remain. Many of the differences originate from the considered linelists - Assumptions regarding linelists, CIAs and isotopic species can lead to notable differences \citep{Fortney2020,Niraula2022}. A detailed report exploring this in detail and for the other ``casks" will be reported soon following this initial protocol report.

\begin{table}
\centering
\caption{List of MALBEC radiative transfer ``casks" and experiments. The exoMIP column describes which partner exoMIP these cases will be coordinated with.}
\begin{tabular}{c c c c c c }
\hline
\hline
Cask & Planet & Geometry & Description & Aerosols & exoMIP \\
\hline
T0A & GJ1214b & Transit & Layering and ray-tracing test & Clear & ~ \\ \hline
T1A & TRAPPIST-1e & Transit & Compact terrestrial atmosphere & Clear & THAI \\ \hline
T1B & TRAPPIST-1e & Transit & Cloudy terrestrial atmosphere & Simple & THAI \\ \hline
T3A & GJ1214b & Transit & Isothermal mini-Neptune & Clear & CAMEMBERT \\ \hline
T3B & GJ1214b & Transit & Extended clear atmosphere & Clear & CAMEMBERT \\ \hline
T3C & GJ1214b & Transit & Cloudy extended atmosphere & Simple & PIE \\ \hline
T4A & \cite{Barstow2020} & Transit & Hot gas-rich atmosphere (model 0) & Clear & MOCHA \\ \hline
T4B & WASP-18b & Transit & Ref. to HST/JWST observations & Clear & MOCHA \\ \hline
D1A & TRAPPIST-1e & Phase 0/90/180 & Compact atmosphere (as T1A) & Clear & THAI \\ \hline
D3B & GJ1214b & Phase 0/90/180 & Deep atmosphere (as T3B) & Clear & CAMEMBERT \\ \hline
D5A & Earth-like & Phase 0/90/180 & Modern Earth atmosphere & Clear & SAMOSA/FILLET\\ \hline
D5B & Earth-like & Phase 0/90/180 & Cloudy modern Earth & Water-ice & CREME \\ \hline
\hline
\hline
\label{tab:cases}
\end{tabular}
\end{table}

\subsection{The transit simulations}

The transit experiments (T0A, T1A, T1B, T3A, T3B, T3C, T4A, T4B) are meant to test the functionalities of the models for the interpretation of transit spectra, which is one of the prime techniques to characterize exoplanets. The resulting spectra in this mode will be primarily defined by the considered opacity processes and the ray-tracing algorithm employed by each model. As such, we have defined a set of experiments to test these models at extreme regimes, highly-compact and dense atmospheres (e.g., TRAPPIST-1 e, \cite{Gillon_2017}), extended and light atmospheres (e.g., GJ1214b, \cite{Charbonneau_2009}), and hot-Jupiters (e.g., \citealt{Barstow2020}). By testing transit simulations at different temperatures we expect to better understand the differences due to various line-list databases (e.g., HITRAN vs. HITEMP vs. ExoMol), while the different ambient compositions (e.g., high molecular weight atmospheres vs. \ce{H2}-rich) would allow us to explore the effects on the assumptions regarding collisional broadening and CIAs. Results from these experiments are primarily tailored to understand the impact of the RT models when interpreting HST, JWST and ARIEL transit data.

As we explore more in detail the origin of the discrepancies of the spectra presented in Figure\ref{Fig:T3B}, linelists and the methods considered to model opacities by each model are the dominating factors. The case of T3B is interesting in that it simulates a hydrogen-rich atmosphere at relatively high temperatures (350-900K), requiring specializing linelists. On the other hand, models primarily developed to model terrestrial planets and temperate planets are set to operate with HITRAN (mostly air collisional parameters and tailored for $<$300K), which is limited in operability for this specific case. When including the millions of transitions active at these high temperatures, correlated-k and pre-computed high-resolution opacity tables are needed, yet how these databases are computed, the source linelists and the way they are handled by each RT model can lead to substantial differences. As users use these models in their default operational regimes, they should be mindful of the limitations of the linelists configured for the simulation. Detailed comparisons between linelists and opacity modeling methods (e.g., line-by-line, correlated-k, and cross-sections) have been discussed elsewhere as previously discussed, and as part of this model intercomparison, we primarily want to identify spectral regions in which models tend to disagree and why. 

\begin{figure}[ht]
\centering
\resizebox{18cm}{!}{\includegraphics{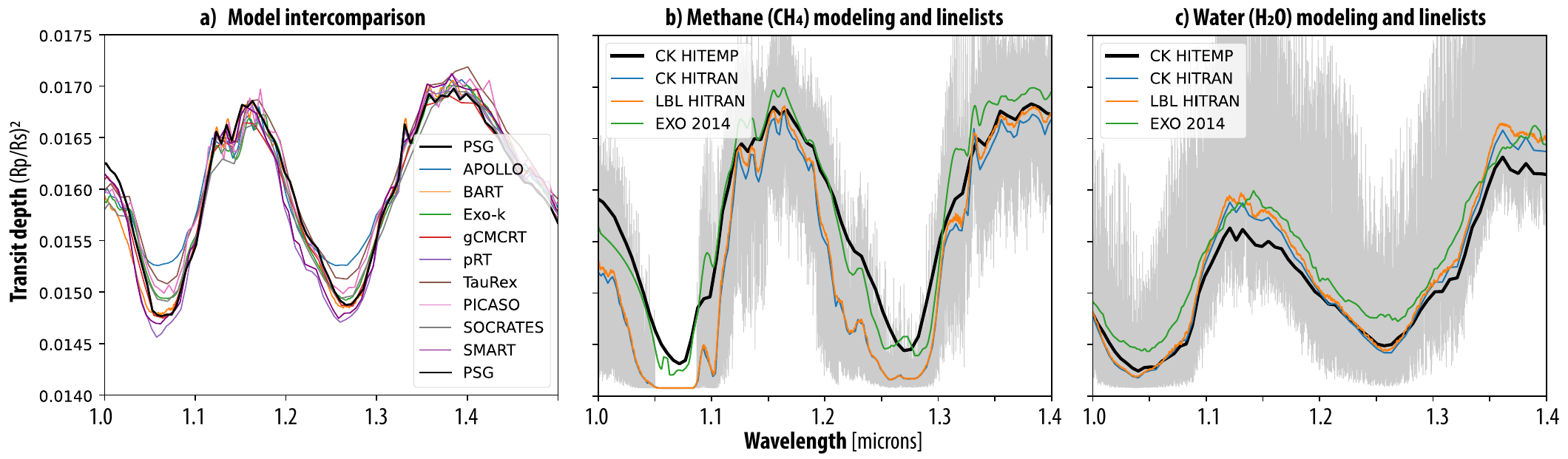}}
\caption{Comparison of model results in the near-infrared and simulations computed employing different linelists (HITRAN, HITEMP and \cite{Freedman2014} [EXO]) and opacity algorithms (line-by-line [LBL] and correlated-k [CK]). There is relatively good agreement in the model inter-comparison, yet notable differences are observed in the valleys at 1.05 and 1.3 $\mu$m regions. Water and methane are the main absorbers at these wavelengths, and how the opacities for these species is treated has a notable effect on how full the valleys look. Panels (b) and (c) shows PSG simulations employing different methods and linelists for methane and water respectively. The thin underlying gray trace is the high-resolution line-by-line spectrum as computed with the HITRAN database, and one can observe substantial differences depending on the choice of opacity database. }
\label{Fig:T3B_LBL}
\end{figure}

In Figure \ref{Fig:T3B_LBL}, we show a zoom of the near infrared region, in which one can see notable differences between the models, and different levels of transit intensity particularly near the ``valley" regions. These valley regions are primarily affected by water and methane opacity effects. In order to demonstrate the effects of linelists in this spectral region of this specific test case, we ran one specific RT model (PSG) but employing different linelists and methods. If measured at extremely high-resolution, the transit spectrum would look substantially spiky as shown with the thin gray trace in panels (b) and (c), yet as we convolve the spectrum to a lower resolution (RP=200), the corresponding spectrum (LBL HITRAN) looks substantially smooth, with small wings and baseline considerations in the line-by-line (LBL) analysis notably affecting the resulting convolved spectrum. The high-energy linelist (HITEMP), which is more suitable for this test case than HITRAN, fills substantially more the valleys though, while the cross sections by \cite{Freedman2014}, which rely on a different set of linelists, show a general better agreement to HITEMP than to HITRAN. The line-by-line (LBL) and correlated-k (CK) methods when considering the same source linelist and resolution should in principle agree perfectly, yet small spectral grid interpolation effects do lead to subtle differences as shown for LBL-HITRAN and CK-HITRAN. 

\begin{figure}[ht]
\centering
\resizebox{18cm}{!}{\includegraphics{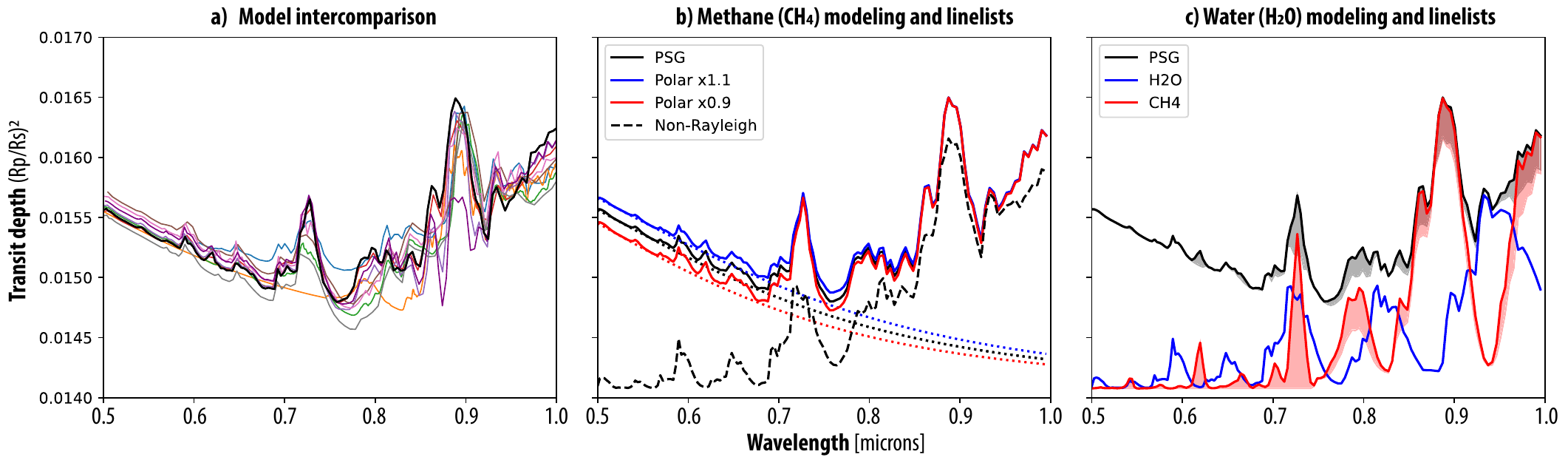}}
\caption{Comparison of preliminary simulation results for test cask T3B. The RT models were run using their typical and core functionalities (e.g., linelists, ray tracing, CIAs) with minimum adaption for this experiment, and were configured to take into account the MALBEC input parameters as close as possible. There is general agreement among the models, with the valleys/peaks in most cases being reproduced by all the models. However, there are specific features in the UV/Optical and at certain wavelengths where the models disagree, which is probably related to the employed linelists and the available CIAs for each RT model. For instance, differences at the near-IR valleys are in most cases related to the considered CH4$_4$ linelists (e.g., HITRAN [less absorption], HITEMP [intermediate], ExoMol [more absorption]). Further revised results with improved parameterizations for each RT model will be reported in a subsequent investigation.}
\label{Fig:T3B_UV}
\end{figure}

A particularly challenging spectral region is the optical and UV, in which linelists normally provide incomplete coverage, and where different RT models treat molecular opacities and Rayleigh scattering in different ways. As such, simulation results differ substantially in this spectral region as the panel (a) in figure \ref{Fig:T3B_UV} shows.  Some RT models (e.g., PSG) compute Rayleigh cross-sections as a summation of the individual molecular cross-sections, and based on the individual polarizability of the encompassing molecules. Subtle differences in how each model handles Rayleigh scattering or the assumed molecular polarizability values would lead to differences in the generated spectra. In panel (b), we perturb the PSG model by scaling the polarizability values by 0.9 and 1.1, and one obtains a level of difference comparable to the observed model differences. More complex is the treatment of molecular opacities at these wavelengths, since only sparse and limited molecular spectroscopy is available. At these high frequencies, the energy structure of molecules become increasingly non-quantized into a broad swath of energy levels. As such, classical linelists do not accurately characterize these high-frequency spectral regions, and laboratory measurements of molecular cross-sections are typically needed to fully capture the opacities at these wavelengths. For instance, the line-by-line and correlated-k simulations in PSG are complemented in this spectral range with laboratory cross-sections collected from a range of spectral databases, with most of these cross-sections originating from the MPI-Mainz Spectral Atlas \citep{keller-rudek_mpi-mainz_2013}. The shaded regions in panel (c) shows how this complements, which is particularly notable for methane near 0.7 $\mu$m. As better and more complete spectroscopy becomes available over time for a wide range of collisional and excitation regimes, the RT models will need to be updated and adapted to ingest these. More details regarding the databases, the methods, and the line-shape considerations employed by each model, and the associated implications and recommendations, will be presented in a subsequent paper that discusses these findings in detail.

\subsection{The direct imaging simulations}

For the direct imaging experiments (D1A, D3B, D5A, D5B), the simulations are to be carried out when the planets are in quadrature (phase 90), the prime phase employed by coronagraphy, and sampling the night-hemisphere (phase 180) to test thermal emissions, and as seen from the star (phase 0, geometric albedo). These modes will allow us to better separate issues when modeling thermal and reflected light, each being susceptible differently to the assumptions regarding multiple-scattering, observational angle, disk discretization and spherical considerations. The simulations are to be carried out employing the highest fidelity available for each model, including multiple-scattering, the highest number of scattering/Legendre terms possible and employing realistic stellar and surface flux models. The tests designed with different levels of complexity. D1A is the simplest test: it has no aerosols, a generally thin atmosphere (with N$_2$, CO$_2$ and H$_2$O starting at 1 atm), and the reflected and thermal emissions are mostly dominated by the modeling of the surface, which is assumed to be simply a Lambert scattering solid surface with a constant albedo and emissivity. Test D3B explores how models handle optically thick atmospheres, in which the deep bottom is defined by opacity functions and not by a solid surface, and it will test how each model handles deep gas atmospheres as those of gas giants. The D3B atmosphere is free of aerosols, so the opacity terms are relatively simple to model and defined by molecular extinction and Rayleigh scattering, yet the treatment of reflected radiation for the quadrature phase will explore how the models discretize and sample the disk across a wide range of illumination regimes in an optically thick multiple scattering non-Lambertian deep atmosphere. The D5A test explores a much more spectrally complex atmosphere of Earth, with strong signatures of H$_2$O, O$_3$, O$_2$, CH$_4$, CO$_2$ and CO, in an ambient air mixture (O$_2$/N$_2$). In particular, this test will be used to analyze the treatment of specialized UV cross-sections and modeling of Rayleigh multiple-scattering across models and RT algorithms. Case D5B is based on D5A but adds the complexity of clouds, and the test is designed to produce a realistic Earth spectrum as targeted by future observatories. One type of cloud is considered in this test, yet the size distribution is assumed to vary across altitude together with its abundance. A specific target of this test is to explore how each model handle handles size distributions of aerosols, optical constants, and multiple-scattering for a wide range of incidence and emission angles.

We expect more differences in the reflected component of the spectrum, in which the effects of multiple-scattering are more notable, and where differences in the treatment of UV opacities and stellar models/templates could become significant. These experiments are specifically tailored to test the differences of RT models in the characterization of large exoplanets with JWST via secondary transit/phase analysis, of gas giants with the Roman Space Telescope, and of Earth-like planets via direct imaging. In particular, future flagship observatories (e.g., Habitable Worlds Observatory, LIFE mission) have as a prime target the characterization of Earth-like planets via direct imaging, so better understanding the current state of RT modeling of these planets is of great importance when designing these future telescopes.

\section{The MALBEC Repositories}
\label{sec:repo}
The input MALBEC files and all the output data resulting from the different simulations and analysis are to be accessible at the MALBEC permanent repository \footnote{\url{https://ckan.emac.gsfc.nasa.gov/organization/cuisines-malbec}} by participating scientists and the exoplanet community. We have prepared the MALBEC input files for each of the experiments and provided preliminary results as computed with at least one of the participating RT models. Ultimately, the results for all models will be made available for public access upon the publication of the results. Pre-publication access can be requested by contacting the authors. Additional inputs and scripts related to the analysis of the data and production of plots for the publications will be made available on the MALBEC GitHub repository \footnote{\url{https://github.com/projectcuisines/malbec}}. 

\section{Summary} 
\label{sec:summary}
In this paper, we presented the protocol for the MALBEC model intercomparison project aiming to compare radiative transfer models used by the exoplanet community. MALBEC is designed to have a low entrance barrier across three sets of experiments, therefore allowing for a broad participation with no requirement to participate in \textit{all} experiments. The sets of experiments allow us to test numerous configurations of exoplanet RT calculations, such as transmission and emission spectroscopy, as well as direct imaging, for terrestrial planets, mini-Neptunes and hot Jupiters. The goal is to highlight any differences arising from different opacity calculations, line broadening, collision-induced-absorption, Rayleigh scattering, layering treatment, etc.
In the context of JWST, and while preparing for future observatories, differences in number of hours of integration to detect molecules of interest will be quantified using the spectra provided by each RT model combined to a single noise model. User-friendly MALBEC ASCII configuration files containing all the required parameters have been created for each of the cases, associated with analytical profiles for some of them.  Output files for at least one RT model are provided to guide the community in the generation of spectra for each experiment. MALBEC will also serve as a bridge between other CUISINES exoMIPs, such as the Photochemical model Intercomparison for Exoplanets (PIE) to provide atmospheric profiles with multiple gas abundances, the TRAPPIST-1 Habitable Atmosphere Intercomparison (THAI) for TRAPPIST-1e clear and cloudy modern Earth-like profiles, and (CAMEMBERT) for GJ1214 b profiles. Participation in MALBEC is open to anyone, with monthly team meetings and annual Building a Unified Framework For Exoplanet Treatments (BUFFET) workshops, with the third edition planned for late 2024.

\section{Acknowledgements}

MALBEC belongs to the CUISINES meta-framework, a Nexus for Exoplanet System Science (NExSS) science working group. GLV, TJF, VK and SF acknowledge support from the GSFC Sellers Exoplanet Environments Collaboration (SEEC), which is funded in part by the NASA Planetary Science Divisions Internal Scientist Funding Model. EA's work has been carried out within the framework of the NCCR PlanetS supported by the Swiss National Science Foundation under grants 51NF40\_182901 and 51NF40\_205606. EL is supported by the SNSF Ambizione Fellowship grant (\#193448). Material produced using Met Office Software. We acknowledge use of the Monsoon2 system, a collaborative facility supplied under the Joint Weather and Climate Research Programme, a strategic partnership between the Met Office and the Natural Environment Research Council. his work was supported by a Leverhulme Trust research project grant [RPG-2020-82], a Science and Technology Facilities Council Consolidated Grant [ST/R000395/1] and a UKRI Future Leaders Fellowship [MR/T040866/1]. MTM \& NJM acknowledge funding from the Bell Burnell Graduate Scholarship Fund, administered and managed by the Institute of Physics, which made this work possible. MDH was supported by the NASA Fellowship Activity under NASA Grant 80NSSC20K0682 and by an appointment to the NASA Postdoctoral Program at the NASA Goddard Space Flight Center, administered by Oak Ridge Associated Universities under contract with NASA. ML and EWS acknowledge support from the Virtual Planetary Laboratory, which is funded via NASA Astrobiology Program Grant No. 80NSSC18K0829, and the the NASA Interdisciplinary Consortia for Astrobiology Research (ICAR) Program via grant Nos. 80NSSC21K0905 and 80NSSC21K0594.
ARH was supported by NASA under award number 80GSFC21M0002 through the CRESST II cooperative agreement. We sincerely thank Duncan Christie for his help with generating SOCRATES spectral files for the T3B case.

\bibliography{biblio}{}

\begin{thebibliography}{}
\expandafter\ifx\csname natexlab\endcsname\relax\def\natexlab#1{#1}\fi
\providecommand{\url}[1]{\href{#1}{#1}}
\providecommand{\dodoi}[1]{doi:~\href{http://doi.org/#1}{\nolinkurl{#1}}}
\providecommand{\doeprint}[1]{\href{http://ascl.net/#1}{\nolinkurl{http://ascl.net/#1}}}
\providecommand{\doarXiv}[1]{\href{https://arxiv.org/abs/#1}{\nolinkurl{https://arxiv.org/abs/#1}}}

\bibitem[{Abel {et~al.}(2011)Abel, Frommhold, Li, \& Hunt}]{Abel2011}
Abel, M., Frommhold, L., Li, X., \& Hunt, K. L.~C. 2011, The Journal of
  Physical Chemistry A, 115, 6805, \dodoi{10.1021/JP109441F}

\bibitem[{{Adams} {et~al.}(2022){Adams}, {Kataria}, {Batalha}, {Gao}, \&
  {Knutson}}]{Adams2022}
{Adams}, D.~J., {Kataria}, T., {Batalha}, N.~E., {Gao}, P., \& {Knutson}, H.~A.
  2022, \apj, 926, 157, \dodoi{10.3847/1538-4357/ac3d32}

\bibitem[{Ahrer {et~al.}(2022)Ahrer, Alderson, Batalha, Batalha, Bean, Beatty,
  Bell, Benneke, Berta-Thompson, Carter, Crossfield, Espinoza, Feinstein,
  Fortney, Gibson, Goyal, Kempton, Kirk, Kreidberg, López-Morales, Line,
  Lothringer, Moran, Mukherjee, Ohno, Parmentier, Piaulet, Rustamkulov,
  Schlawin, Sing, Stevenson, Wakeford, Allen, Birkmann, Brande, Crouzet,
  Cubillos, Damiano, Désert, Gao, Harrington, Hu, Kendrew, Knutson, Lagage,
  Leconte, Lendl, MacDonald, May, Miguel, Molaverdikhani, Moses, Murray,
  Nehring, Nikolov, Petit dit de~la Roche, Radica, Roy, Stassun, Taylor,
  Waalkes, Wachiraphan, Welbanks, Wheatley, Aggarwal, Alam, Banerjee, Barstow,
  Blecic, Casewell, Changeat, Chubb, Colón, Coulombe, Daylan, de~Val-Borro,
  Decin, Dos~Santos, Flagg, France, Fu, García~Muñoz, Gizis, Glidden, Grant,
  Heng, Henning, Hong, Inglis, Iro, Kataria, Komacek, Krick, Lee, Lewis,
  Lillo-Box, Lustig-Yaeger, Mancini, Mandell, Mansfield, Marley, Mikal-Evans,
  Morello, Nixon, Ortiz~Ceballos, Piette, Powell, Rackham, Ramos-Rosado,
  Rauscher, Redfield, Rogers, Roman, Roudier, Scarsdale, Shkolnik, Southworth,
  Spake, Steinrueck, Tan, Teske, Tremblin, Tsai, Tucker, Turner, Valenti,
  Venot, Waldmann, Wallack, Zhang, Zieba, \& {JWST Transiting Exoplanet
  Community Early Release Science Team}}]{ahrer2022}
Ahrer, E.-M., Alderson, L., Batalha, N.~M., {et~al.} 2022, Nature, 1,
  \dodoi{10.1038/s41586-022-05269-w}

\bibitem[{Al-Refaie {et~al.}(2021)Al-Refaie, Changeat, Venot, Waldmann, \&
  Tinetti}]{Al-Refaie2021A3.1}
Al-Refaie, A.~F., Changeat, Q., Venot, O., Waldmann, I.~P., \& Tinetti, G.
  2021, \apj

\bibitem[{{Alei} {et~al.}(2022){Alei}, {Konrad}, {Angerhausen}, {Grenfell},
  {Molli{\`e}re}, {Quanz}, {Rugheimer}, {Wunderlich}, \& {the LIFE
  collaboration}}]{Alei2022}
{Alei}, E., {Konrad}, B.~S., {Angerhausen}, D., {et~al.} 2022, arXiv e-prints,
  arXiv:2204.10041.
\newblock \doarXiv{2204.10041}

\bibitem[{{Amundsen} {et~al.}(2014){Amundsen}, {Baraffe}, {Tremblin},
  {Manners}, {Hayek}, {Mayne}, \& {Acreman}}]{Amundsen2014}
{Amundsen}, D.~S., {Baraffe}, I., {Tremblin}, P., {et~al.} 2014, \aap, 564,
  A59, \dodoi{10.1051/0004-6361/201323169}

\bibitem[{{Amundsen} {et~al.}(2016){Amundsen}, {Mayne}, {Baraffe}, {Manners},
  {Tremblin}, {Drummond}, {Smith}, {Acreman}, \& {Homeier}}]{Amundsen2016}
{Amundsen}, D.~S., {Mayne}, N.~J., {Baraffe}, I., {et~al.} 2016, \aap, 595,
  A36, \dodoi{10.1051/0004-6361/201629183}

\bibitem[{Arney {et~al.}(2014)Arney, Meadows, Crisp, Schmidt, Bailey, \&
  Robinson}]{Arney2014}
Arney, G., Meadows, V., Crisp, D., {et~al.} 2014, J. Geophys. Res. Planets,
  119, 1860, \dodoi{10.1002/2014je004662}

\bibitem[{{Arney} {et~al.}(2017){Arney}, {Meadows}, {Domagal-Goldman},
  {Deming}, {Robinson}, {Tovar}, {Wolf}, \& {Schwieterman}}]{Arney2017}
{Arney}, G.~N., {Meadows}, V.~S., {Domagal-Goldman}, S.~D., {et~al.} 2017,
  \apj, 836, 49, \dodoi{10.3847/1538-4357/836/1/49}

\bibitem[{Barstow {et~al.}(2022)Barstow, Changeat, Chubb, Cubillos, Edwards,
  MacDonald, Min, \& Waldmann}]{Barstow2022}
Barstow, J.~K., Changeat, Q., Chubb, K.~L., {et~al.} 2022, Experimental
  Astronomy, 53, 447, \dodoi{10.1007/s10686-021-09821-w}

\bibitem[{{Barstow} {et~al.}(2020){Barstow}, {Changeat}, {Garland}, {Line},
  {Rocchetto}, \& {Waldmann}}]{Barstow2020}
{Barstow}, J.~K., {Changeat}, Q., {Garland}, R., {et~al.} 2020, \mnras, 493,
  4884, \dodoi{10.1093/mnras/staa548}

\bibitem[{{Batalha} \& {Rooney}(2020)}]{Batalha2020}
{Batalha}, N., \& {Rooney}, C. 2020, {natashabatalha/picaso: Release 2.1}, 2.1,
  Zenodo,  Zenodo, \dodoi{10.5281/zenodo.4206648}

\bibitem[{{Batalha} {et~al.}(2021){Batalha}, {Rooney}, \&
  {MacDonald}}]{Batalha2021}
{Batalha}, N., {Rooney}, C., \& {MacDonald}, R. 2021, {natashabatalha/picaso:
  Release 2.2}, v2.2.0, Zenodo,  Zenodo, \dodoi{10.5281/zenodo.5093710}

\bibitem[{{Batalha} {et~al.}(2019){Batalha}, {Marley}, {Lewis}, \&
  {Fortney}}]{Batalha2019}
{Batalha}, N.~E., {Marley}, M.~S., {Lewis}, N.~K., \& {Fortney}, J.~J. 2019,
  \apj, 878, 70, \dodoi{10.3847/1538-4357/ab1b51}

\bibitem[{Baudino {et~al.}(2017)Baudino, Molli{\`{e}}re, Venot, Tremblin,
  B{\'{e}}zard, \& Lagage}]{Baudino2017}
Baudino, J.-L., Molli{\`{e}}re, P., Venot, O., {et~al.} 2017, The Astrophysical
  Journal, 850, 150, \dodoi{10.3847/1538-4357/aa95be}

\bibitem[{Blecic {et~al.}(2022)Blecic, Harrington, Cubillos, Bowman, Rojo,
  Stemm, Challener, Himes, Foster, Dobbs-Dixon, Foster, Lust, Blumenthal,
  Bruce, \& Loredo}]{BlecicEtal2022psjBART3}
Blecic, J., Harrington, J., Cubillos, P.~E., {et~al.} 2022, The Planetary
  Science Journal, 3, 82, \dodoi{10.3847/psj/ac3515}

\bibitem[{{Borysow}(2002)}]{Borysow2002aapH2H2CIA}
{Borysow}, A. 2002, \aap, 390, 779, \dodoi{10.1051/0004-6361:20020555}

\bibitem[{{Borysow} {et~al.}(2001){Borysow}, {Jorgensen}, \&
  {Fu}}]{Borysow2001jqsrtH2H2CIA}
{Borysow}, A., {Jorgensen}, U.~G., \& {Fu}, Y. 2001, \jqsrt, 68, 235,
  \dodoi{10.1016/S0022-4073(00)00023-6}

\bibitem[{{Cahoy} {et~al.}(2010){Cahoy}, {Marley}, \& {Fortney}}]{Cahoy2010}
{Cahoy}, K.~L., {Marley}, M.~S., \& {Fortney}, J.~J. 2010, \apj, 724, 189,
  \dodoi{10.1088/0004-637X/724/1/189}

\bibitem[{Caldas {et~al.}(2019)Caldas, Leconte, Selsis, Waldmann, Bord\'e,
  Rocchetto, \& Charnay}]{Caldas2019}
Caldas, A., Leconte, J., Selsis, F., {et~al.} 2019, A\&A, 623, A161,
  \dodoi{10.1051/0004-6361/201834384}

\bibitem[{Charbonneau {et~al.}(2009)Charbonneau, Berta, Irwin, Burke, Nutzman,
  Buchhave, Lovis, Bonfils, Latham, Udry, Murray-Clay, Holman, Falco, Winn,
  Queloz, Pepe, Mayor, Delfosse, \& Forveille}]{Charbonneau_2009}
Charbonneau, D., Berta, Z.~K., Irwin, J., {et~al.} 2009, Nature, 462, 891,
  \dodoi{10.1038/nature08679}

\bibitem[{Charnay {et~al.}(2015)Charnay, Meadows, Misra, Leconte, \&
  Arney}]{Charnay2015}
Charnay, B., Meadows, V., Misra, A., Leconte, J., \& Arney, G. 2015, The
  Astrophysical Journal, 813, L1, \dodoi{10.1088/2041-8205/813/1/L1}

\bibitem[{{Chaverot} {et~al.}(2022){Chaverot}, {Turbet}, {Bolmont}, \&
  {Leconte}}]{CTB22}
{Chaverot}, G., {Turbet}, M., {Bolmont}, E., \& {Leconte}, J. 2022, \aap, 658,
  A40, \dodoi{10.1051/0004-6361/202142286}

\bibitem[{Christie {et~al.}(2022)Christie, Lee, Innes, Noti, Charnay, Fauchez,
  Mayne, Deitrick, Ding, Greco, Hammond, Malsky, Mandell, Rauscher, Roman,
  Sergeev, Sohl, Steinrueck, Turbet, Wolf, Zamyatina, \&
  Carone}]{christie_camembert_2022}
Christie, D.~A., Lee, E. K.~H., Innes, H., {et~al.} 2022, The Planetary Science
  Journal, 3, 261, \dodoi{10.3847/PSJ/ac9dfe}

\bibitem[{Chubb {et~al.}(2021)Chubb, Rocchetto, Yurchenko, Min, Waldmann,
  Barstow, Mollière, Al-Refaie, Phillips, \& Tennyson}]{Chubb2021}
Chubb, K.~L., Rocchetto, M., Yurchenko, S.~N., {et~al.} 2021, Astronomy \&
  Astrophysics, 646, A21, \dodoi{10.1051/0004-6361/202038350}

\bibitem[{Cox(2015)}]{Cox2015}
Cox, A. 2015, Allen’s Astrophysical Quantities (Springer).
\newblock
  \url{https://books.google.com/books?hl=en&lr=&id=TjDtCAAAQBAJ&oi=fnd&pg=PR5&dq=cox+2015+physical+quantities&ots=QVL64iCyLR&sig=_UqV0ooARxJtdwT7eK7u_FN2IZE#v=onepage&q=cox%202015%20physical%20quantities&f=false}

\bibitem[{Crisp(1997)}]{Crisp1997}
Crisp, D. 1997, Geophysical Research Letters, 24, 571,
  \dodoi{10.1029/97GL50245}

\bibitem[{{{Cubillos}} {et~al.}(2017){{Cubillos}}, {{Harrington}}, {Loredo},
  {{Lust}}, {{Blecic}}, \& {{Stemm}}}]{CubillosEtal2017apjRednoise}
{{Cubillos}}, P., {{Harrington}}, J., {Loredo}, T.~J., {et~al.} 2017, \aj, 153,
  3, \dodoi{10.3847/1538-3881/153/1/3}

\bibitem[{{Cubillos}(2017)}]{Cubillos2017apjRepack}
{Cubillos}, P.~E. 2017, \apj, 850, 32, \dodoi{10.3847/1538-4357/aa9228}

\bibitem[{Cubillos \& Blecic(2021)}]{Cubillos2021}
Cubillos, P.~E., \& Blecic, J. 2021, Monthly Notices of the Royal Astronomical
  Society, 505, 2675, \dodoi{10.1093/mnras/stab1405}

\bibitem[{Cubillos {et~al.}(2022)Cubillos, Harrington, Blecic, Himes, Rojo,
  Loredo, Lust, Challener, Foster, Stemm, Foster, \&
  Blumenthal}]{CubillosEtal2022psjBART2}
Cubillos, P.~E., Harrington, J., Blecic, J., {et~al.} 2022, The Planetary
  Science Journal, 3, 81, \dodoi{10.3847/psj/ac348b}

\bibitem[{Dahlback \& Stamnes(1991)}]{dahlback1991}
Dahlback, A., \& Stamnes, K. 1991, Planetary and Space Science, 39, 671,
  \dodoi{10.1016/0032-0633(91)90061-E}

\bibitem[{Deitrick {et~al.}(2023)Deitrick, Haqq-Misra, Kadoya, Ramirez,
  Simonetti, Barnes, \& Fauchez}]{Deitrick2023}
Deitrick, R., Haqq-Misra, J., Kadoya, S., {et~al.} 2023, Functionality of Ice
  Line Latitudinal EBM Tenacity (FILLET). Protocol Version 1.0. A CUISINES
  intercomparison project,  arXiv, \dodoi{10.48550/ARXIV.2302.04980}

\bibitem[{Edwards \& Slingo(1996)}]{Edwards1996}
Edwards, J.~M., \& Slingo, A. 1996, Quarterly Journal of the Royal
  Meteorological Society, 122, 689, \dodoi{10.1002/qj.49712253107}

\bibitem[{{Eyring} {et~al.}(2016){Eyring}, {Bony}, {Meehl}, {Senior},
  {Stevens}, {Stouffer}, \& {Taylor}}]{cmip6}
{Eyring}, V., {Bony}, S., {Meehl}, G.~A., {et~al.} 2016, Geoscientific Model
  Development, 9, 1937, \dodoi{10.5194/gmd-9-1937-2016}

\bibitem[{{Falco} {et~al.}(2022){Falco}, {Zingales}, {Pluriel}, \&
  {Leconte}}]{FZP22}
{Falco}, A., {Zingales}, T., {Pluriel}, W., \& {Leconte}, J. 2022, \aap, 658,
  A41, \dodoi{10.1051/0004-6361/202141940}

\bibitem[{Fally {et~al.}(2000)Fally, Vandaele, Carleer, Hermans, Jenouvrier,
  Mérienne, Coquart, \& Colin}]{fally2000}
Fally, S., Vandaele, A.~C., Carleer, M., {et~al.} 2000, Journal of Molecular
  Spectroscopy, 204, 10, \dodoi{10.1006/jmsp.2000.8204}

\bibitem[{{Fauchez} {et~al.}(2019){Fauchez}, {Turbet}, {Villanueva}, {Wolf},
  {Arney}, {Kopparapu}, {Lincowski}, {Mandell}, {de Wit}, {Pidhorodetska},
  {Domagal-Goldman}, \& {Stevenson}}]{Fauchez2019}
{Fauchez}, T.~J., {Turbet}, M., {Villanueva}, G.~L., {et~al.} 2019, \apj, 887,
  194, \dodoi{10.3847/1538-4357/ab5862}

\bibitem[{Fauchez {et~al.}(2020)Fauchez, Villanueva, Schwieterman, Turbet,
  Arney, Pidhorodetska, Kopparapu, Mandell, \& Domagal-Goldman}]{fauchez2020}
Fauchez, T.~J., Villanueva, G.~L., Schwieterman, E.~W., {et~al.} 2020, Nature
  Astronomy, 4, 372, \dodoi{10.1038/s41550-019-0977-7}

\bibitem[{Fauchez {et~al.}(2022)Fauchez, Villanueva, Sergeev, Turbet, Boutle,
  Tsigaridis, Way, Wolf, Domagal-Goldman, Forget, Haqq-Misra, Kopparapu,
  Manners, \& Mayne}]{fauchez_thai}
Fauchez, T.~J., Villanueva, G.~L., Sergeev, D.~E., {et~al.} 2022, The Planetary
  Science Journal, 3, 213, \dodoi{10.3847/PSJ/ac6cf1}

\bibitem[{Fletcher {et~al.}(2018)Fletcher, Gustafsson, \& Orton}]{Fletcher2018}
Fletcher, L.~N., Gustafsson, M., \& Orton, G.~S. 2018, The Astrophysical
  Journal Supplement Series, 235, 24, \dodoi{10.3847/1538-4365/AAA07A}

\bibitem[{{Foote} {et~al.}(2022){Foote}, {Lewis}, {Kilpatrick}, {Goyal},
  {Bruno}, {Wakeford}, {Robbins-Blanch}, {Kataria}, {MacDonald},
  {L{\'o}pez-Morales}, {Sing}, {Mikal-Evans}, {Bourrier}, {Henry}, \&
  {Buchhave}}]{Foote2022}
{Foote}, T.~O., {Lewis}, N.~K., {Kilpatrick}, B.~M., {et~al.} 2022, \aj, 163,
  7, \dodoi{10.3847/1538-3881/ac2f4a}

\bibitem[{{Foreman-Mackey} {et~al.}(2013){Foreman-Mackey}, {Hogg}, {Lang}, \&
  {Goodman}}]{Foreman-Mackey2013}
{Foreman-Mackey}, D., {Hogg}, D.~W., {Lang}, D., \& {Goodman}, J. 2013, \pasp,
  125, 306, \dodoi{10.1086/670067}

\bibitem[{{Fortney} {et~al.}(2020){Fortney}, {Visscher}, {Marley}, {Hood},
  {Line}, {Thorngren}, {Freedman}, \& {Lupu}}]{Fortney2020}
{Fortney}, J.~J., {Visscher}, C., {Marley}, M.~S., {et~al.} 2020, \aj, 160,
  288, \dodoi{10.3847/1538-3881/abc5bd}

\bibitem[{{Freedman} {et~al.}(2014){Freedman}, {Lustig-Yaeger}, {Fortney},
  {Lupu}, {Marley}, \& {Lodders}}]{Freedman2014}
{Freedman}, R.~S., {Lustig-Yaeger}, J., {Fortney}, J.~J., {et~al.} 2014, \apjs,
  214, 25, \dodoi{10.1088/0067-0049/214/2/25}

\bibitem[{{Garland} \& {Irwin}(2019)}]{garland2019}
{Garland}, R., \& {Irwin}, P.~G.~J. 2019, arXiv e-prints, arXiv:1903.03997,
  \dodoi{10.48550/arXiv.1903.03997}

\bibitem[{{Gharib-Nezhad} {et~al.}(2021){Gharib-Nezhad}, {Iyer}, {Line},
  {Freedman}, {Marley}, \& {Batalha}}]{GaribNezhad2021}
{Gharib-Nezhad}, E., {Iyer}, A., {Line}, M.~R., {et~al.} 2021, in 2021
  International Symposium on Molecular Spectroscopy,
  \dodoi{10.15278/isms.2021.WF04}

\bibitem[{Gillon {et~al.}(2017)Gillon, Triaud, Demory, Jehin, Agol, Deck,
  Lederer, de~Wit, Burdanov, Ingalls, Bolmont, Leconte, Raymond, Selsis,
  Turbet, Barkaoui, Burgasser, Burleigh, Carey, Chaushev, Copperwheat, Delrez,
  Fernandes, Holdsworth, Kotze, Grootel, Almleaky, Benkhaldoun, Magain, \&
  Queloz}]{Gillon_2017}
Gillon, M., Triaud, A. H. M.~J., Demory, B.-O., {et~al.} 2017, Nature, 542,
  456, \dodoi{10.1038/nature21360}

\bibitem[{Goldblatt {et~al.}(2017)Goldblatt, Kavanagh, \&
  Dewey}]{Goldblatt2017}
Goldblatt, C., Kavanagh, L., \& Dewey, M. 2017, Geoscientific Model
  Development, 10, 3931, \dodoi{10.5194/gmd-10-3931-2017}

\bibitem[{Gordon {et~al.}(2017)Gordon, Rothman, Hill, Kochanov, Tan, Bernath,
  Birk, Boudon, Campargue, Chance, Drouin, Flaud, Gamache, Hodges, Jacquemart,
  Perevalov, Perrin, Shine, Smith, Tennyson, Toon, Tran, Tyuterev, Barbe,
  Császár, Devi, Furtenbacher, Harrison, Hartmann, Jolly, Johnson, Karman,
  Kleiner, Kyuberis, Loos, Lyulin, Massie, Mikhailenko, Moazzen-Ahmadi,
  Müller, Naumenko, Nikitin, Polyansky, Rey, Rotger, Sharpe, Sung, Starikova,
  Tashkun, Auwera, Wagner, Wilzewski, Wcisło, Yu, \& Zak}]{Gordon2017}
Gordon, I.~E., Rothman, L.~S., Hill, C., {et~al.} 2017, Journal of Quantitative
  Spectroscopy and Radiative Transfer, 203, 3,
  \dodoi{10.1016/J.JQSRT.2017.06.038}

\bibitem[{{Gordon} {et~al.}(2022){Gordon}, {Rothman}, {Hargreaves}, {Hashemi},
  {Karlovets}, {Skinner}, {Conway}, {Hill}, {Kochanov}, {Tan}, {Wcis{\l}o},
  {Finenko}, {Nelson}, {Bernath}, {Birk}, {Boudon}, {Campargue}, {Chance},
  {Coustenis}, {Drouin}, {Flaud}, {Gamache}, {Hodges}, {Jacquemart}, {Mlawer},
  {Nikitin}, {Perevalov}, {Rotger}, {Tennyson}, {Toon}, {Tran}, {Tyuterev},
  {Adkins}, {Baker}, {Barbe}, {Can{\`e}}, {Cs{\'a}sz{\'a}r}, {Dudaryonok},
  {Egorov}, {Fleisher}, {Fleurbaey}, {Foltynowicz}, {Furtenbacher}, {Harrison},
  {Hartmann}, {Horneman}, {Huang}, {Karman}, {Karns}, {Kassi}, {Kleiner},
  {Kofman}, {Kwabia-Tchana}, {Lavrentieva}, {Lee}, {Long}, {Lukashevskaya},
  {Lyulin}, {Makhnev}, {Matt}, {Massie}, {Melosso}, {Mikhailenko}, {Mondelain},
  {M{\"u}ller}, {Naumenko}, {Perrin}, {Polyansky}, {Raddaoui}, {Raston},
  {Reed}, {Rey}, {Richard}, {T{\'o}bi{\'a}s}, {Sadiek}, {Schwenke},
  {Starikova}, {Sung}, {Tamassia}, {Tashkun}, {Vander Auwera}, {Vasilenko},
  {Vigasin}, {Villanueva}, {Vispoel}, {Wagner}, {Yachmenev}, \&
  {Yurchenko}}]{GordonEtal2022jqsrtHITRAN2020}
{Gordon}, I.~E., {Rothman}, L.~S., {Hargreaves}, R.~J., {et~al.} 2022, Journal
  of Quantitative Spectroscopy and Radiative Transfer, 277, 107949,
  \dodoi{10.1016/j.jqsrt.2021.107949}

\bibitem[{{Goyal} {et~al.}(2018){Goyal}, {Mayne}, {Sing}, {Drummond},
  {Tremblin}, {Amundsen}, {Evans}, {Carter}, {Spake}, {Baraffe}, {Nikolov},
  {Manners}, {Chabrier}, \& {Hebrard}}]{Goyal2018}
{Goyal}, J.~M., {Mayne}, N., {Sing}, D.~K., {et~al.} 2018, \mnras, 474, 5158,
  \dodoi{10.1093/mnras/stx3015}

\bibitem[{Grimm {et~al.}(2021)Grimm, Malik, Kitzmann, Guzmán-Mesa,
  Hoeijmakers, Fisher, Mendonça, Yurchenko, Tennyson, Alesina, Buchschacher,
  Burnier, Segransan, Kurucz, \& Heng}]{Grimm_2021}
Grimm, S.~L., Malik, M., Kitzmann, D., {et~al.} 2021, The Astrophysical Journal
  Supplement Series, 253, 30, \dodoi{10.3847/1538-4365/abd773}

\bibitem[{Haqq-Misra {et~al.}(2022)Haqq-Misra, Wolf, Fauchez, Shields, \&
  Kopparapu}]{HQsamosa}
Haqq-Misra, J., Wolf, E.~T., Fauchez, T.~J., Shields, A.~L., \& Kopparapu,
  R.~K. 2022, The Planetary Science Journal, 3, 260, \dodoi{10.3847/PSJ/ac9479}

\bibitem[{{Hargreaves} {et~al.}(2020){Hargreaves}, {Gordon}, {Rey}, {Nikitin},
  {Tyuterev}, {Kochanov}, \& {Rothman}}]{HargreavesEtal2020apjsMethaneHITEMP}
{Hargreaves}, R.~J., {Gordon}, I.~E., {Rey}, M., {et~al.} 2020, \apjs, 247, 55,
  \dodoi{10.3847/1538-4365/ab7a1a}

\bibitem[{{Hargreaves} {et~al.}(2019){Hargreaves}, {Gordon}, {Rothman},
  {Tashkun}, {Perevalov}, {Lukashevskaya}, {Yurchenko}, {Tennyson}, \&
  {M{\"u}ller}}]{Hargreaves2019}
{Hargreaves}, R.~J., {Gordon}, I.~E., {Rothman}, L.~S., {et~al.} 2019, \jqsrt,
  232, 35, \dodoi{10.1016/j.jqsrt.2019.04.040}

\bibitem[{Harrington {et~al.}(2022)Harrington, Himes, Cubillos, Blecic, Rojo,
  Challener, Lust, Bowman, Blumenthal, Dobbs-Dixon, Foster, Foster, Green,
  Loredo, McIntyre, Stemm, \& Wright}]{HarringtonEtal2022psjBART1}
Harrington, J., Himes, M.~D., Cubillos, P.~E., {et~al.} 2022, The Planetary
  Science Journal, 3, 80, \dodoi{10.3847/psj/ac3513}

\bibitem[{{Himes} \& {Harrington}(2022)}]{HimesHarrington2022apjWASP12b}
{Himes}, M.~D., \& {Harrington}, J. 2022, \apj, 931, 86,
  \dodoi{10.3847/1538-4357/ac1e9f}

\bibitem[{Hogan \& Matricardi(2020)}]{Hogan_Matricardi2020}
Hogan, R.~J., \& Matricardi, M. 2020, Geoscientific Model Development, 13,
  6501, \dodoi{10.5194/gmd-13-6501-2020}

\bibitem[{{Howe} {et~al.}(2022){Howe}, {McElwain}, \& {Mandell}}]{Howe2022}
{Howe}, A.~R., {McElwain}, M.~W., \& {Mandell}, A.~M. 2022, \apj, 935, 107,
  \dodoi{10.3847/1538-4357/ac5590}

\bibitem[{Irwin {et~al.}(2008)Irwin, Teanby, {de Kok}, Fletcher, Howett, Tsang,
  Wilson, Calcutt, Nixon, \& Parrish}]{Irwin2008}
Irwin, P., Teanby, N., {de Kok}, R., {et~al.} 2008, Journal of Quantitative
  Spectroscopy and Radiative Transfer, 109, 1136,
  \dodoi{https://doi.org/10.1016/j.jqsrt.2007.11.006}

\bibitem[{{Karman} {et~al.}(2019){Karman}, {Gordon}, {van der Avoird},
  {Baranov}, {Boulet}, {Drouin}, {Groenenboom}, {Gustafsson}, {Hartmann},
  {Kurucz}, {Rothman}, {Sun}, {Sung}, {Thalman}, {Tran}, {Wishnow},
  {Wordsworth}, {Vigasin}, {Volkamer}, \& {van der Zande}}]{2019Karman}
{Karman}, T., {Gordon}, I.~E., {van der Avoird}, A., {et~al.} 2019, \icarus,
  328, 160, \dodoi{10.1016/j.icarus.2019.02.034}

\bibitem[{Keller-Rudek {et~al.}(2013)Keller-Rudek, Moortgat, Sander, \&
  Sörensen}]{keller-rudek_mpi-mainz_2013}
Keller-Rudek, H., Moortgat, G.~K., Sander, R., \& Sörensen, R. 2013, Earth
  System Science Data, 5, 365, \dodoi{10.5194/essd-5-365-2013}

\bibitem[{Kofman \& Villanueva(2021)}]{kofman_absorption_2021}
Kofman, V., \& Villanueva, G.~L. 2021, Journal of Quantitative Spectroscopy and
  Radiative Transfer, 270, 107708,
  \dodoi{https://doi.org/10.1016/j.jqsrt.2021.107708}

\bibitem[{{Konrad} {et~al.}(2022){Konrad}, {Alei}, {Quanz}, {Angerhausen},
  {Carri{\'o}n-Gonz{\'a}lez}, {Fortney}, {Grenfell}, {Kitzmann},
  {Molli{\`e}re}, {Rugheimer}, {Wunderlich}, \& {LIFE
  Collaboration}}]{Konrad2022}
{Konrad}, B.~S., {Alei}, E., {Quanz}, S.~P., {et~al.} 2022, \aap, 664, A23,
  \dodoi{10.1051/0004-6361/202141964}

\bibitem[{{Kozakis} {et~al.}(2022){Kozakis}, {Mendon{\c{c}}a}, \&
  {Buchhave}}]{Kozakis2022}
{Kozakis}, T., {Mendon{\c{c}}a}, J.~M., \& {Buchhave}, L.~A. 2022, \aap, 665,
  A156, \dodoi{10.1051/0004-6361/202244164}

\bibitem[{{Leconte}(2021)}]{2021A&A...645A..20L}
{Leconte}, J. 2021, \aap, 645, A20, \dodoi{10.1051/0004-6361/202039040}

\bibitem[{Lee {et~al.}(2019)Lee, Taylor, Grimm, Baudino, Garland, Irwin, \&
  Wood}]{Lee2019}
Lee, E., Taylor, J., Grimm, S.~L., {et~al.} 2019, Monthly Notices of the Royal
  Astronomical Society, 487, 2082, \dodoi{10.1093/mnras/stz1418}

\bibitem[{{Lee} {et~al.}(2022){Lee}, {Wardenier}, {Prinoth}, {Parmentier},
  {Grimm}, {Baeyens}, {Carone}, {Christie}, {Deitrick}, {Kitzmann}, {Mayne},
  {Roman}, \& {Thorsbro}}]{Lee2022}
{Lee}, E. K.~H., {Wardenier}, J.~P., {Prinoth}, B., {et~al.} 2022, \apj, 929,
  180, \dodoi{10.3847/1538-4357/ac61d6}

\bibitem[{{Lew} {et~al.}(2022){Lew}, {Apai}, {Zhou}, {Marley}, {Mayorga},
  {Tan}, {Parmentier}, {Casewell}, \& {Xu (许偲艺)}}]{Lew2022}
{Lew}, B. W.~P., {Apai}, D., {Zhou}, Y., {et~al.} 2022, \aj, 163, 8,
  \dodoi{10.3847/1538-3881/ac3001}

\bibitem[{{Li} {et~al.}(2015){Li}, {Gordon}, {Rothman}, {Tan}, {Hu}, {Kassi},
  {Campargue}, \& {Medvedev}}]{Li2015}
{Li}, G., {Gordon}, I.~E., {Rothman}, L.~S., {et~al.} 2015, \apjs, 216, 15,
  \dodoi{10.1088/0067-0049/216/1/15}

\bibitem[{Lincowski {et~al.}(2018)Lincowski, Meadows, Crisp, Robinson, Luger,
  Lustig-Yaeger, \& Arney}]{Lincowski2018}
Lincowski, A.~P., Meadows, V.~S., Crisp, D., {et~al.} 2018, Astrophys. J., 867,
  76, \dodoi{10.3847/1538-4357/aae36a}

\bibitem[{Lines {et~al.}(2018)Lines, Manners, Mayne, Goyal, Carter, Boutle,
  Lee, Helling, Drummond, Acreman, \& Sing}]{Lines18_exonephology}
Lines, S., Manners, J., Mayne, N.~J., {et~al.} 2018, Monthly Notices of the
  Royal Astronomical Society, 481, 194, \dodoi{10.1093/mnras/sty2275}

\bibitem[{{Lines} {et~al.}(2018){Lines}, {Manners}, {Mayne}, {Goyal}, {Carter},
  {Boutle}, {Lee}, {Helling}, {Drummond}, {Acreman}, \& {Sing}}]{Lines2018}
{Lines}, S., {Manners}, J., {Mayne}, N.~J., {et~al.} 2018, \mnras, 481, 194,
  \dodoi{10.1093/mnras/sty2275}

\bibitem[{Lustig-Yaeger {et~al.}(2019)Lustig-Yaeger, Meadows, \&
  Lincowski}]{Lustig-Yaeger2019}
Lustig-Yaeger, J., Meadows, V.~S., \& Lincowski, A.~P. 2019, The Astronomical
  Journal, 158, 27, \dodoi{10.3847/1538-3881/ab21e0}

\bibitem[{MacDonald \& Madhusudhan(2017)}]{MacDonald2017}
MacDonald, R.~J., \& Madhusudhan, N. 2017, Monthly Notices of the Royal
  Astronomical Society, 469, 1979, \dodoi{10.1093/mnras/stx804}

\bibitem[{Mai \& Line(2019)}]{MaiLine2019}
Mai, C., \& Line, M.~R. 2019, The Astrophysical Journal, 883, 144,
  \dodoi{10.3847/1538-4357/ab3e6d}

\bibitem[{{Malik} {et~al.}(2017){Malik}, {Grosheintz}, {Mendon{\c{c}}a},
  {Grimm}, {Lavie}, {Kitzmann}, {Tsai}, {Burrows}, {Kreidberg}, {Bedell},
  {Bean}, {Stevenson}, \& {Heng}}]{Malik_2017}
{Malik}, M., {Grosheintz}, L., {Mendon{\c{c}}a}, J.~M., {et~al.} 2017, \aj,
  153, 56, \dodoi{10.3847/1538-3881/153/2/56}

\bibitem[{Manners {et~al.}(2022)Manners, Edwards, Hill, \& Thelen}]{Manners22}
Manners, J., Edwards, J.~M., Hill, P., \& Thelen, J.-C. 2022, SOCRATES ({S}uite
  {O}f {C}ommunity {RA}diative {T}ransfer codes based on {E}dwards and
  {S}lingo) Technical Guide, Met Office, UK.
\newblock \url{https://code.metoffice.gov.uk/trac/socrates}

\bibitem[{{Marley} {et~al.}(1999){Marley}, {Gelino}, {Stephens}, {Lunine}, \&
  {Freedman}}]{Marley1999}
{Marley}, M.~S., {Gelino}, C., {Stephens}, D., {Lunine}, J.~I., \& {Freedman},
  R. 1999, \apj, 513, 879, \dodoi{10.1086/306881}

\bibitem[{{McKay} {et~al.}(1989){McKay}, {Pollack}, \& {Courtin}}]{McKay1989}
{McKay}, C.~P., {Pollack}, J.~B., \& {Courtin}, R. 1989, \icarus, 80, 23,
  \dodoi{10.1016/0019-1035(89)90160-7}

\bibitem[{Meador \& Weaver(1980)}]{Meador_Weaver_1980}
Meador, W.~E., \& Weaver, W.~R. 1980, Journal of Atmospheric Sciences, 37, 630
  , \dodoi{https://doi.org/10.1175/1520-0469(1980)037<0630:TSATRT>2.0.CO;2}

\bibitem[{Meadows \& Crisp(1996)}]{MeadowsCrisp1996}
Meadows, V.~S., \& Crisp, D. 1996, Journal of Geophysical Research: Planets,
  101, 4595, \dodoi{https://doi.org/10.1029/95JE03567}

\bibitem[{Min {et~al.}(2020)Min, Ormel, Chubb, Helling, \& Kawashima}]{Min2020}
Min, M., Ormel, C.~W., Chubb, K., Helling, C., \& Kawashima, Y. 2020, Astronomy
  {\&} Astrophysics, 642, A28, \dodoi{10.1051/0004-6361/201937377}

\bibitem[{{Mlawer} {et~al.}(2012){Mlawer}, {Payne}, {Moncet}, {Delamere},
  {Alvarado}, \& {Tobin}}]{2012Mlawer}
{Mlawer}, E.~J., {Payne}, V.~H., {Moncet}, J.~L., {et~al.} 2012, Philosophical
  Transactions of the Royal Society of London Series A, 370, 2520,
  \dodoi{10.1098/rsta.2011.0295}

\bibitem[{{Molli{\`e}re} {et~al.}(2020){Molli{\`e}re}, {Stolker}, {Lacour},
  {Otten}, {Shangguan}, {Charnay}, {Molyarova}, {Nowak}, {Henning}, {Marleau},
  {Semenov}, {van Dishoeck}, {Eisenhauer}, {Garcia}, {Garcia Lopez}, {Girard},
  {Greenbaum}, {Hinkley}, {Kervella}, {Kreidberg}, {Maire}, {Nasedkin},
  {Pueyo}, {Snellen}, {Vigan}, {Wang}, {de Zeeuw}, \& {Zurlo}}]{Molliere2020}
{Molli{\`e}re}, P., {Stolker}, T., {Lacour}, S., {et~al.} 2020, \aap, 640,
  A131, \dodoi{10.1051/0004-6361/202038325}

\bibitem[{{Molli\`ere, P.} {et~al.}(2019){Molli\`ere, P.}, {Wardenier, J. P.},
  {van Boekel, R.}, {Henning, Th.}, {Molaverdikhani, K.}, \& {Snellen, I. A.
  G.}}]{Molliere2019}
{Molli\`ere, P.}, {Wardenier, J. P.}, {van Boekel, R.}, {et~al.} 2019, A\&A,
  627, A67, \dodoi{10.1051/0004-6361/201935470}

\bibitem[{{Mukherjee} {et~al.}(2023){Mukherjee}, {Batalha}, {Fortney}, \&
  {Marley}}]{Mukherjee2023}
{Mukherjee}, S., {Batalha}, N.~E., {Fortney}, J.~J., \& {Marley}, M.~S. 2023,
  \apj, 942, 71, \dodoi{10.3847/1538-4357/ac9f48}

\bibitem[{{Mukherjee} {et~al.}(2021){Mukherjee}, {Fortney}, {Jensen-Clem},
  {Tan}, {Marley}, \& {Batalha}}]{Mukherjee2021}
{Mukherjee}, S., {Fortney}, J.~J., {Jensen-Clem}, R., {et~al.} 2021, \apj, 923,
  113, \dodoi{10.3847/1538-4357/ac2d92}

\bibitem[{Mérienne {et~al.}(2001)Mérienne, Jenouvrier, Coquart, Carleer,
  Fally, Colin, Vandaele, \& Hermans}]{merienne2001}
Mérienne, M.-F., Jenouvrier, A., Coquart, B., {et~al.} 2001, Journal of
  Molecular Spectroscopy, 207, 120, \dodoi{10.1006/jmsp.2001.8314}

\bibitem[{Niraula {et~al.}(2022)Niraula, de~Wit, Gordon, Hargreaves,
  Sousa-Silva, \& Kochanov}]{Niraula2022}
Niraula, P., de~Wit, J., Gordon, I.~E., {et~al.} 2022, Nature Astronomy, 6,
  1287, \dodoi{10.1038/s41550-022-01773-1}

\bibitem[{Paradise {et~al.}(2022)Paradise, Macdonald, Menou, Lee, \&
  Fan}]{paradise_exoplasim_2022}
Paradise, A., Macdonald, E., Menou, K., Lee, C., \& Fan, B.~L. 2022, Monthly
  Notices of the Royal Astronomical Society, 511, 3272,
  \dodoi{10.1093/mnras/stac172}

\bibitem[{{Partridge} \&
  {Schwenke}(1997)}]{PartridgeSchwenke1997jcpLineListH2O}
{Partridge}, H., \& {Schwenke}, D.~W. 1997, \jcp, 106, 4618,
  \dodoi{10.1063/1.473987}

\bibitem[{Pincus {et~al.}(2020)Pincus, Buehler, Brath, Crevoisier, Jamil,
  Franklin~Evans, Manners, Menzel, Mlawer, Paynter, Pernak, \&
  Tellier}]{Pincus2020}
Pincus, R., Buehler, S.~A., Brath, M., {et~al.} 2020, Journal of Geophysical
  Research: Atmospheres, 125, e2020JD033483,
  \dodoi{https://doi.org/10.1029/2020JD033483}

\bibitem[{Pinty {et~al.}(2001)Pinty, Gobron, Widlowski, Gerstl, Verstraete,
  Antunes, Bacour, Gascon, Gastellu, Goel, Jacquemoud, North, Qin, \&
  Thompson}]{Pinty2001}
Pinty, B., Gobron, N., Widlowski, J.-L., {et~al.} 2001, Journal of Geophysical
  Research: Atmospheres, 106, 11937,
  \dodoi{https://doi.org/10.1029/2000JD900493}

\bibitem[{Polyansky {et~al.}(2018)Polyansky, Kyuberis, Zobov, Tennyson,
  Yurchenko, \& Lodi}]{polyanski}
Polyansky, O.~L., Kyuberis, A.~A., Zobov, N.~F., {et~al.} 2018, Monthly Notices
  of the Royal Astronomical Society, 480, 2597, \dodoi{10.1093/mnras/sty1877}

\bibitem[{{Robbins-Blanch} {et~al.}(2022){Robbins-Blanch}, {Kataria},
  {Batalha}, \& {Adams}}]{Robbins-Blanch2022}
{Robbins-Blanch}, N., {Kataria}, T., {Batalha}, N.~E., \& {Adams}, D.~J. 2022,
  \apj, 930, 93, \dodoi{10.3847/1538-4357/ac658c}

\bibitem[{Robinson(2017)}]{Robinson2017}
Robinson, T.~D. 2017, The Astrophysical Journal, 836, 236,
  \dodoi{10.3847/1538-4357/aa5ea8}

\bibitem[{Robinson {et~al.}(2011)Robinson, Meadows, Crisp, Deming, A'hearn,
  Charbonneau, Livengood, Seager, Barry, Hearty, Hewagama, Lisse, McFadden, \&
  Wellnitz}]{Robinson2011}
Robinson, T.~D., Meadows, V.~S., Crisp, D., {et~al.} 2011, Astrobiology, 11,
  393, \dodoi{10.1089/ast.2011.0642}

\bibitem[{{Rojo}(2006)}]{Rojo2006PhD}
{Rojo}, P.~M. 2006, PhD thesis, Cornell University

\bibitem[{{Rooney} {et~al.}(2022){Rooney}, {Batalha}, {Gao}, \&
  {Marley}}]{Rooney2022}
{Rooney}, C.~M., {Batalha}, N.~E., {Gao}, P., \& {Marley}, M.~S. 2022, \apj,
  925, 33, \dodoi{10.3847/1538-4357/ac307a}

\bibitem[{Rothman {et~al.}(2014)Rothman, Gordon, Rothman, \&
  Gordon}]{Rothman2014}
Rothman, L.~S., Gordon, I.~E., Rothman, L.~S., \& Gordon, I.~E. 2014, hitr, 49,
  \dodoi{10.5281/ZENODO.11207}

\bibitem[{{Rothman} {et~al.}(2010){Rothman}, {Gordon}, {Barber}, {Dothe},
  {Gamache}, {Goldman}, {Perevalov}, {Tashkun}, \&
  {Tennyson}}]{RothmanEtal2010jqsrtHITEMP2010}
{Rothman}, L.~S., {Gordon}, I.~E., {Barber}, R.~J., {et~al.} 2010, \jqsrt, 111,
  2139, \dodoi{10.1016/j.jqsrt.2010.05.001}

\bibitem[{{Schwenke}(1998)}]{Schwenke1998fadiLineListTiO}
{Schwenke}, D.~W. 1998, Faraday Discussions, 109, 321, \dodoi{10.1039/a800070k}

\bibitem[{Serdyuchenko {et~al.}(2014)Serdyuchenko, Gorshelev, Weber, Chehade,
  \& Burrows}]{serdyuchenko2014}
Serdyuchenko, A., Gorshelev, V., Weber, M., Chehade, W., \& Burrows, J.~P.
  2014, Atmos. Meas. Tech., 7, 625, \dodoi{10.5194/amt-7-625-2014}

\bibitem[{Sergeev {et~al.}(2022)Sergeev, Fauchez, Turbet, Boutle, Tsigaridis,
  Way, Wolf, Domagal-Goldman, Forget, Haqq-Misra, Kopparapu, Lambert, Manners,
  \& Mayne}]{sergeev_thai}
Sergeev, D.~E., Fauchez, T.~J., Turbet, M., {et~al.} 2022, The Planetary
  Science Journal, 3, 212, \dodoi{10.3847/PSJ/ac6cf2}

\bibitem[{{Sneep} \& {Ubachs}(2005)}]{2005Sneep_Ubachs}
{Sneep}, M., \& {Ubachs}, W. 2005, \jqsrt, 92, 293,
  \dodoi{10.1016/j.jqsrt.2004.07.025}

\bibitem[{Stamnes {et~al.}(2000)Stamnes, Tsay, \& Laszlo}]{stamnes2000}
Stamnes, K., Tsay, S.-C., \& Laszlo, I. 2000, NASA Technical Report

\bibitem[{Stamnes {et~al.}(1988)Stamnes, Tsay, Wiscombe, \&
  Jayaweera}]{Stamnes1988}
Stamnes, K., Tsay, S.~C., Wiscombe, W., \& Jayaweera, K. 1988, Applied optics,
  27, 2502, \dodoi{10.1364/AO.27.002502}

\bibitem[{Tennyson {et~al.}(2016)Tennyson, Yurchenko, Al-Refaie, Barton, Chubb,
  Coles, Diamantopoulou, Gorman, Hill, Lam, Lodi, McKemmish, Na, Owens,
  Polyansky, Rivlin, Sousa-Silva, Underwood, Yachmenev, \& Zak}]{Tennyson2016}
Tennyson, J., Yurchenko, S.~N., Al-Refaie, A.~F., {et~al.} 2016, Journal of
  Molecular Spectroscopy, 327, 73, \dodoi{10.1016/J.JMS.2016.05.002}

\bibitem[{{The JWST Transiting Exoplanet Community Early Release Science Team}
  {et~al.}(2022){The JWST Transiting Exoplanet Community Early Release Science
  Team}, {Ahrer}, {Alderson}, {Batalha}, {Batalha}, {Bean}, {Beatty}, {Bell},
  {Benneke}, {Berta-Thompson}, {Carter}, {Crossfield}, {Espinoza}, {Feinstein},
  {Fortney}, {Gibson}, {Goyal}, {Kempton}, {Kirk}, {Kreidberg},
  {L{\'o}pez-Morales}, {Line}, {Lothringer}, {Moran}, {Mukherjee}, {Ohno},
  {Parmentier}, {Piaulet}, {Rustamkulov}, {Schlawin}, {Sing}, {Stevenson},
  {Wakeford}, {Allen}, {Birkmann}, {Brande}, {Crouzet}, {Cubillos}, {Damiano},
  {D{\'e}sert}, {Gao}, {Harrington}, {Hu}, {Kendrew}, {Knutson}, {Lagage},
  {Leconte}, {Lendl}, {MacDonald}, {May}, {Miguel}, {Molaverdikhani}, {Moses},
  {Murray}, {Nehring}, {Nikolov}, {Petit dit de la Roche}, {Radica}, {Roy},
  {Stassun}, {Taylor}, {Waalkes}, {Wachiraphan}, {Welbanks}, {Wheatley},
  {Aggarwal}, {Alam}, {Banerjee}, {Barstow}, {Blecic}, {Casewell}, {Changeat},
  {Chubb}, {Col{\'o}n}, {Coulombe}, {Daylan}, {de Val-Borro}, {Decin}, {Dos
  Santos}, {Flagg}, {France}, {Fu}, {Garc{\'\i}a Mu{\~n}oz}, {Gizis},
  {Glidden}, {Grant}, {Heng}, {Henning}, {Hong}, {Inglis}, {Iro}, {Kataria},
  {Komacek}, {Krick}, {Lee}, {Lewis}, {Lillo-Box}, {Lustig-Yaeger}, {Mancini},
  {Mandell}, {Mansfield}, {Marley}, {Mikal-Evans}, {Morello}, {Nixon}, {Ortiz
  Ceballos}, {Piette}, {Powell}, {Rackham}, {Ramos-Rosado}, {Rauscher},
  {Redfield}, {Rogers}, {Roman}, {Roudier}, {Scarsdale}, {Shkolnik},
  {Southworth}, {Spake}, {E Steinrueck}, {Tan}, {Teske}, {Tremblin}, {Tsai},
  {Tucker}, {Turner}, {Valenti}, {Venot}, {Waldmann}, {Wallack}, {Zhang}, \&
  {Zieba}}]{JWSTers}
{The JWST Transiting Exoplanet Community Early Release Science Team}, {Ahrer},
  E.-M., {Alderson}, L., {et~al.} 2022, arXiv e-prints, arXiv:2208.11692,
  \dodoi{10.48550/arXiv.2208.11692}

\bibitem[{Tinetti {et~al.}(2005)Tinetti, Meadows, Crisp, Fong, Velusamy, \&
  Snively}]{Tinetti2005}
Tinetti, G., Meadows, V.~S., Crisp, D., {et~al.} 2005, Astrobiology, 5, 461,
  \dodoi{10.1089/ast.2005.5.461}

\bibitem[{{Toon} {et~al.}(1989{\natexlab{a}}){Toon}, {McKay}, {Ackerman}, \&
  {Santhanam}}]{Toon1989}
{Toon}, O.~B., {McKay}, C.~P., {Ackerman}, T.~P., \& {Santhanam}, K.
  1989{\natexlab{a}}, \jgr, 94, 16287, \dodoi{10.1029/JD094iD13p16287}

\bibitem[{{Toon} {et~al.}(1989{\natexlab{b}}){Toon}, {McKay}, {Ackerman}, \&
  {Santhanam}}]{TMA89}
---. 1989{\natexlab{b}}, \jgr, 94, 16287, \dodoi{10.1029/JD094iD13p16287}

\bibitem[{{Toon} {et~al.}(1977){Toon}, {Pollack}, \& {Sagan}}]{Toon1977}
{Toon}, O.~B., {Pollack}, J.~B., \& {Sagan}, C. 1977, \icarus, 30, 663,
  \dodoi{10.1016/0019-1035(77)90088-4}

\bibitem[{Turbet {et~al.}(2022)Turbet, Fauchez, Sergeev, Boutle, Tsigaridis,
  Way, Wolf, Domagal-Goldman, Forget, Haqq-Misra, Kopparapu, Lambert, Manners,
  Mayne, \& Sohl}]{turbet_thai}
Turbet, M., Fauchez, T.~J., Sergeev, D.~E., {et~al.} 2022, The Planetary
  Science Journal, 3, 211, \dodoi{10.3847/PSJ/ac6cf0}

\bibitem[{Venot {et~al.}(2018)Venot, Bénilan, Fray, Gazeau, Lefèvre,
  Es-sebbar, Hébrard, Schwell, Bahrini, Montmessin, Lefèvre, \&
  Waldmann}]{venot2018}
Venot, O., Bénilan, Y., Fray, N., {et~al.} 2018, Astronomy \& Astrophysics,
  609, A34, \dodoi{10.1051/0004-6361/201731295}

\bibitem[{{Villanueva} {et~al.}(2022){Villanueva}, {Liuzzi}, {Faggi},
  {Protopapa}, {Kofman}, {Fauchez}, {Stone}, \& {Mandell}}]{Villanueva2022}
{Villanueva}, G.~L., {Liuzzi}, G., {Faggi}, S., {et~al.} 2022, {Fundamentals of
  the Planetary Spectrum Generator} (NASA Goddard Space Flight Center)

\bibitem[{{Villanueva} {et~al.}(2018){Villanueva}, {Smith}, {Protopapa},
  {Faggi}, \& {Mandell}}]{Villanueva2018}
{Villanueva}, G.~L., {Smith}, M.~D., {Protopapa}, S., {Faggi}, S., \&
  {Mandell}, A.~M. 2018, Journal of Quantitative Spectroscopy and Radiative
  Transfer, 217, 86, \dodoi{10.1016/j.jqsrt.2018.05.023}

\bibitem[{Waldmann {et~al.}(2015b)Waldmann, Rocchetto, Tinetti, Barton,
  Yurchenko, \& Tennyson}]{Waldmann2015b}
Waldmann, I.~P., Rocchetto, M., Tinetti, G., {et~al.} 2015b, The Astrophysical
  Journal, 813, 13, \dodoi{10.1088/0004-637X/813/1/13}

\bibitem[{Waldmann {et~al.}(2015)Waldmann, Tinetti, Rocchetto, Barton,
  Yurchenko, \& Tennyson}]{Waldmann2015}
Waldmann, I.~P., Tinetti, G., Rocchetto, M., {et~al.} 2015, The Astrophysical
  Journal, 802, 107, \dodoi{10.1088/0004-637x/802/2/107}

\bibitem[{Waldmann {et~al.}(2015a)Waldmann, Tinetti, Rocchetto, Barton,
  Yurchenko, \& Tennyson}]{Waldmann2015a}
---. 2015a, The Astrophysical Journal, 802, 107,
  \dodoi{10.1088/0004-637X/802/2/107}

\bibitem[{Wolf {et~al.}(2022)Wolf, Kopparapu, Haqq-Misra, \&
  Fauchez}]{wolf_exocam_2022}
Wolf, E.~T., Kopparapu, R., Haqq-Misra, J., \& Fauchez, T.~J. 2022, The
  Planetary Science Journal, 3, 7, \dodoi{10.3847/PSJ/ac3f3d}

\bibitem[{Wunderlich {et~al.}(2020)Wunderlich, Scheucher, Godolt, Grenfell,
  Schreier, Schneider, Wilson, Sánchez-López, López-Puertas, \&
  Rauer}]{Wunderlich2020}
Wunderlich, F., Scheucher, M., Godolt, M., {et~al.} 2020, The Astrophysical
  Journal, 901, 126, \dodoi{10.3847/1538-4357/aba59c}

\bibitem[{Yang {et~al.}(2016)Yang, Leconte, Wolf, Goldblatt, Feldl, Merlis,
  Wang, Koll, Ding, Forget, \& Abbot}]{Yang2016}
Yang, J., Leconte, J., Wolf, E.~T., {et~al.} 2016, The Astrophysical Journal,
  826, 222, \dodoi{10.3847/0004-637X/826/2/222}

\bibitem[{Yurchenko {et~al.}(2017)Yurchenko, Amundsen, Tennyson, \&
  Waldmann}]{Yurchenko2017}
Yurchenko, S.~N., Amundsen, D.~S., Tennyson, J., \& Waldmann, I.~P. 2017,
  Astronomy \& Astrophysics, 605, A95, \dodoi{10.1051/0004-6361/201731026}

\bibitem[{Zawada {et~al.}(2021)Zawada, Franssens, Loughman, Mikkonen, Rozanov,
  Emde, Bourassa, Dueck, Lindqvist, Ramon, Rozanov, Dekemper, Kyr\"ol\"a,
  Burrows, Fussen, \& Degenstein}]{Zawada_amt2021}
Zawada, D., Franssens, G., Loughman, R., {et~al.} 2021, Atmospheric Measurement
  Techniques, 14, 3953, \dodoi{10.5194/amt-14-3953-2021}

\end{thebibliography}
\bibliographystyle{aasjournal}
\end{document}